\documentclass[11pt,aps,amsmath,amssymb,nofootinbib,notitlepage,longbibliography]{revtex4-1}
\usepackage{amsmath,amssymb}
\usepackage{slashed}
\usepackage{bbold}
\usepackage{comment}
\usepackage{graphicx}
\usepackage{bm}
\usepackage[top=1.0in,bottom=1.0in,left=1.0in,right=1.0in]{geometry}
\usepackage[colorlinks,linkcolor=blue,citecolor=blue]{hyperref}
\usepackage{subfig}
\usepackage{array}

\newcolumntype{P}[1]{>{\centering\arraybackslash}p{#1}}

\newcommand{\be}{\begin{equation}}
	\newcommand{\ee}{\end{equation}}
\newcommand{\bea}{\begin{eqnarray}}
	\newcommand{\eea}{\end{eqnarray}}
\newcommand{\ba}{\begin{eqnarray}}
	\newcommand{\ea}{\end{eqnarray}}

\begin{document}

	\title{Pion and Kaon Distribution Amplitudes up to twist-3\\ in the QCD Instanton Vacuum}

	\author{Arthur Kock}
	\email{arthur.kock@stonybrook.edu}
	\affiliation{Center for Nuclear Theory, Department of Physics and Astronomy, Stony Brook University, Stony Brook, New York 11794--3800, USA}

	\author{ Ismail Zahed}
	\email{ismail.zahed@stonybrook.edu}
	\affiliation{Center for Nuclear Theory, Department of Physics and Astronomy, Stony Brook University, Stony Brook, New York 11794--3800, USA}

	%\author{
	%Yizhuang Liu and Ismail Zahed }
	
	%\affiliation{ Department of Physics and Astronomy, \\ Stony Brook University,\\
	%Stony Brook, NY 11794, USA}
	
	\begin{abstract}
		We discuss the pion and kaon  distribution amplitudes up to twist-3 in the context of the random instanton vacuum (RIV).
		We construct explicitly the pertinent quasi-pion and quasi-kaon distributions in the RIV, and analyze them in leading order in the diluteness factor,
		at a resolution fixed by the inverse instanton size. The distribution amplitudes (DA) follow from the large momentum limit.
		The results at higher resolution are discussed using QCD evolution, and compared to their asymptotic limits and some lattice results.
	\end{abstract}
	
	\maketitle
	
	%%%%%%%%%%%%%%%%%%%%%%%%%%%%%%%%%%%%%%%%
	\section{Introduction}

	Light cone distributions are central to the description of hard inclusive and exclusive processes. Thanks to factorization,
	a hard process factors  into a perturbatively calculable contribution times pertinent parton distribution and fragmentation
	functions. Standard examples can be found in deep inelastic scattering, Drell-Yan process and jet production to cite a few.

	The parton distribution functions are defined on the light front, and their moments usually fitted using large empirical data banks.  They
	are not readily amenable  to a non-perturbative and first principle formulation using lattice simulations.
	This situation has by now changed. Ji~\cite{Ji:2013dva} has put forth the concept of space-like quasi-parton distributions that  are perturbatively
	matched to the time-like light-cone distributions~\cite{Zhang:2017bzy,Ji:2015qla,Bali:2018spj,Alexandrou:2018eet,Izubuchi:2019lyk,Izubuchi:2018srq}. 
	This conjecture can be checked to hold non-perturbatively in two-dimensional QCD at next-to-leading order
	in the large $N_c$ limit~\cite{Ji:2018waw}. The quasi-parton distribution matrix elements calculated in a fixed size Euclidean lattice QCD, have been 
	argued to  match those obtained through LSZ reduction in continuum Minkowski QCD, to all orders in perturbation theory~\cite{Briceno2017}. Some  variants of this
	formulation can be found in the form of pseudo distributions~\cite{Radyushkin:2017gjd}, and lattice cross sections~\cite{Ma:2014jla}.
	A  number of QCD lattice collaborations have implemented some of these ideas, with some reasonable success in extracting
	the light cone parton distributions.
	
	A good understanding of the non-perturbative gauge fields responsible for chiral symmetry breaking 
	was achieved in the context of the QCD instanton vacuum. 
	Several QCD lattice simulations have shown  that the bulk characteristics and correlations  in the QCD vacuum
	are mostly unaffected by lattice cooling~\cite{Chu:1994vi} where quantum effects are pruned,  suggesting that semi-classical gauge
	and fermionic fields dominate  the ground state structure. At weak coupling, instantons and anti-instantons
	are exact semi-classical gauge tunneling configurations with large actions and finite topological charge which
	support exact quark zero modes with specific chirality. They are at the origin of the spontaneous breaking of chiral
	symmetry and the emergence of a hadronic mass for the low-lying hadronic excitations such as the pion, kaon and nucleon. 
	Orbitally excited hadrons are more sensitive to confinement, perhaps in the extended QCD instanton-dyon vacuum~\cite{Diakonov:2007nv,Liu:2015ufa},
	or in the QCD instanton vacuum with long P-vortices~\cite{Greensite:2016pfc,Biddle:2019gke}.
	
	In this work we follow up on our recent study of the quasi-distributions in the random QCD instanton 
	vacuum (RIV)~\cite{Kock:2020}. More specifically, we will analyze the two-particle pion and kaon quasi-distributions up to twist-3 in the RIV, and
	extract the light cone distribution amplitudes in the large momentum limit. The moments of the twist-3 pion  distribution 
	amplitudes in an effective model of the RIV, and the twist-3 pion distribution amplitudes in a light front quark model using light cone signature,
	were recently discussed in~\cite{Nam:2006mb,Choi:2016meb}. Since the RIV vacuum is Euclidean, the distribution amplitudes are naturally
	extracted from the quasi-distributions with space-like signature.

	%Since quasi-distributions calculated using Euclidean signature 
	%produce the same light-cone distributions post-matching, the RIV, which is more naturally formulated in Euclidean signature, is appropriate 
	%for the task of calculating  quasi-distributions~\cite{Ji:2013dva}.
	
	The outline of the paper is as follows: In section~\ref{sec_instantons} we briefly review the salient features of the RIV. In section~\ref{twist-3-general}
	we discuss the general structure of the pion and kaon  in terms of the twist-2 and twist-3 contributions. Although the latters are subleading  at asymptotic
	momenta in say the pion electromagnetic form factor, they still contribute substantially in the pre-asymptotic regime. In section~\ref{QDA-3} we define
	the quasi-pion   and quasi-kaon distribution amplitudes and analyze them in the RIV using the power counting in the diluteness factor detailed in~\cite{Kock:2020}.
	The massless and massive pseudoscalar and pseudotensor twist-3 pion and kaon   distribution amplitudes are then extracted in the large momentum limit at the
	resolution fixed by the instanton size. The  twist-2,-3 pion and kaon 
	distribution amplitudes at higher resolution are discussed in section~\ref{evolution} using the  ERBL  evolution and compared to their asymptotic limits and 
	some lattice results.
	Our conclusions are in section~\ref{sec_conclusions}.  Some useful details are found in the appendices.

	\begin{figure}[h!]
		\begin{center}
			\includegraphics[width=16cm]{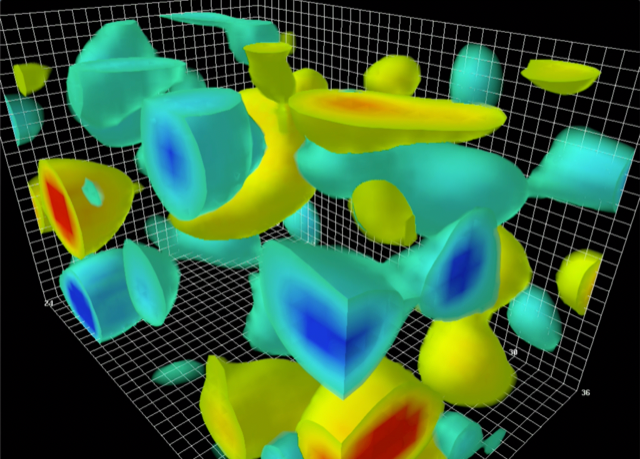}
			\caption{Instantons (yellow) and anti-instantons (blue) configurations in the cooled YM vacuum~\cite{Moran:2008xq}.
			}
			\label{fig_VAC}
		\end{center}
	\end{figure}

	\section{Instanton effects}
	\label{sec_instantons}
	
	The cooled QCD vacuum is populated with strong and inhomogeneous topological gauge configurations, i.e.
	instantons and anti-instantons as illustrated in Fig.~\ref{fig_VAC}
	The bulk characteristics  of this vacuum were predicted long ago~\cite{Shuryak:1981ff}

	\begin{equation}
		n_{I+\bar I}\approx 1\, {\rm fm}^{-4}, \,\, \,\,\, \rho\sim \frac 13 \, {\rm fm} \sim \frac 1{0.6} \, {\rm GeV}^{-1}  \label{eqn_ILM}
	\end{equation}
	for the instanton plus anti-instanton density and size, respectively. They combine in the dimensionless parameter
	
	$$\kappa\equiv \pi^2\rho^4 n_{I+\bar I}\approx 3.186\times 10^{-3}$$
	a measure of  the diluteness of the instanton-anti-instanton ensemble in the QCD vacuum. Previous
	lattice simulations using cooling methods support these observations - see \cite{Schafer:1996wv} for a review.
	
	Instanton fields are strong, since their field strengths  are  large.
	For the dominant  size instantons  with $\rho\approx 0.30\, {\rm fm}$ typical for chiral symmetry breaking,
	the fields are very strong at the center
	
	%\begin{equation}
	%\big( G^a_{\mu\nu}(x) \big)^2 = {192 \rho^4 \over (x^2+\rho^2)^4}
	%\end{equation}
	%and typically of order
	
	$$\sqrt{G^2_{\mu\nu}(0)}=\sqrt{192}/\rho^2\approx 5 \, {\rm GeV}^2$$
	Their scale is comparable to the matching scale in the hard and perturbative matching kernels~\cite{Ji:2020ect} 
	which may suggest non-perturbative improvements~\cite{Liu:2021evw}. Their
	contribution can be assessed using semi-classics.   The size distribution of the instantons
	and anti-instantons in the QCD vacuum is well captured semi-empirically by~\cite{Hasenfratz:1999ng,Shuryak:1999fe}
	
	\begin{equation}
		\label{dn_dist}
		dn(\rho) \sim  {d\rho \over \rho^5}\big(\rho \Lambda_{QCD} \big)^{b_{QCD}} \, e^{-\alpha^\prime \rho^2}
	\end{equation}
	with $b_{QCD}=11N_c/3-2N_f/3\approx 9$ (one loop) and $\alpha^\prime =1/2m_\rho^2$
	(rho meson  slope).

	\section{Twist and  chiral structures of the DA of the pion}~\label{twist-3-general}

	In the QCD instanton vacuum, the pion  DA is captured by the vertex
	$\pi^-(p)\rightarrow d_{if\alpha}(k) u^{\dagger}_{jg\beta}(k-p)$, which  corresponds formally to the connected amplitude

	\begin{eqnarray}
		\label{WF1}
		&&\int_{-\infty}^{+\infty}\frac{p^+dz^-}{2\pi}e^{-ixp\cdot z}\left<0\left|\overline{u}_\beta(0)[0,z]d_\alpha(z)\right|\pi^-(p)\right>\nonumber\\
		&&=
		\left(+\frac{if_\pi}4 \gamma^5\bigg(\slashed{p}\,\phi^A_{\pi^-}(x)
		-\chi_\pi \phi_{\pi^-}^P(x)+i \chi_\pi
		\sigma_{\mu\nu}\frac{p^\mu p^{\prime\nu}}{p\cdot p^\prime}  {\phi_{\pi^-}^{\prime\,T}(x)\over {6}}\bigg)\right)_{\alpha\beta}
	\end{eqnarray}
	and its conjugate
	
	\begin{eqnarray}
		\label{WF1X1}
		&&\int_{-\infty}^{+\infty}\frac{p^{\prime -}dz^+}{2\pi}e^{ixp^\prime\cdot z}\left<\pi^-(p^\prime)\left|\overline{d}_\beta(z)[z,0]u_\alpha(0)\right|0\right>\nonumber\\
		&&=
		\left(-\frac{if_\pi}4 \gamma^5\bigg(\slashed{p}^\prime\,\phi^A_{\pi^-}(x)
		+\chi_\pi \phi_{\pi^-}^P(x)-i \chi_\pi
		\sigma_{\mu\nu}\frac{p^\mu p^{\prime\nu}}{p\cdot p^\prime}  {\phi_{\pi^-}^{\prime\,T}(x)\over {6}}\bigg)\right)_{\alpha\beta}
	\end{eqnarray}
up to twist-3.  $[x,y]$  refers to the gauge link, $\sigma_{\mu\nu}=\frac i2[\gamma_\mu, \gamma_\nu]$, $\{\alpha,\beta\}$ represent spinor indices,
	and $\phi^{T\prime}_\pi(x)=\partial_x \phi^{T}_\pi(x)$.
	(\ref{WF1}-\ref{WF1X1}) are explicitly odd under P parity. Note that the 4-vector $p'_\mu$ appears  in the DA of a pion with 4-vector $p_\mu$, in reference to the conjugate
	light-cone direction, with  generally  no relation to the second pion.
	In the DA of a  pion with momentum $p'_\mu$, the exchange  $p\leftrightarrow p^\prime$ needs to be enforced,
	effectively flipping the sign of the last term.
	
	In (\ref{WF1}), (\ref{WF1X1}), and subsequent derivations, the ket $|\pi^-(P)\rangle$ refers to the physical negative-pion state. 
	(Note the switch in flavors if the current $J_{\pi^-}(x)=\bar{d}(x)i\gamma^5 u(x)$ is used to define the pion state).
	(\ref{WF1}-\ref{WF1X1})  can be  inverted, to recast the pion twist-2 and twist-3 light-cone wavefunctions in  explicit form

	\begin{subequations}\label{MinkDAs}
	\begin{eqnarray}
		&&\phi_{\pi^-}^A(x)=
		\frac  1{if_\pi}\int_{-\infty} ^{+\infty} \frac{dz^-}{2\pi}e^{ixp\cdot z}\left<0\left|\overline{u}(0)\gamma^+\gamma_5[0,z]d(z)\right|\pi^-(p)\right>\\
		&&\phi^P_{\pi^-}(x)=
		\frac  {p^+}{f_\pi\chi_\pi}\int_{-\infty} ^{+\infty}  \frac{dz^-}{2\pi}e^{ixp\cdot z}\left<0\left|\overline{u}(0)i\gamma_5[0,z]d(z)\right|\pi^-(p)\right>\\
		&&{\phi^{T\prime}_{\pi^-}(x)}=
		\frac  6{f_\pi\chi_{\pi}}\frac {p^\mu p^{\prime \nu}p^+}{p\cdot p^\prime}\int _{-\infty} ^{+\infty} \frac{dz^-}{2\pi}e^{ixp\cdot z}\left<0\left|\overline{u}(0)\sigma_{\mu\nu}\gamma_5[0,z]d(z)\right|\pi^-(p)\right>
	\end{eqnarray}
	\end{subequations}
with  all DAs normalized to 1. The prime in the last relation refers to $\partial_x\phi_{\pi^-}^T (x)$.
	The leading twist-2 DA  $\phi^A_\pi(x)$ is chirally-diagonal.  Its normalization to 1 is fixed by 
	the weak  pion decay constant $f_\pi\approx 130\,{\rm MeV}$,

	\begin{equation}
		\label{WF11}
		\left<0\left|\overline{u}(0)\gamma^\mu{(1-\gamma^5)}d(0)\right|\pi^-(p)\right>=
		-{\rm Tr}\bigg(\gamma^\mu{(1-\gamma^5)}\bigg(\frac{if_\pi}4\gamma^5\slashed{p}\bigg)\bigg)\,\int_0^1dx\,\phi^A_{\pi^-}(x)\equiv 
		if_\pi\,p^\mu
	\end{equation}
	%Here $\phi_{\pi^-}(x)$, the pion light front distribution, is normalized to 1.
	Isospin symmetry and charge conjugation force
	$\phi_\pi(x)=\phi_\pi(\overline x)$.
	%v2
	The  two twist-3  independent DAs 
	$\phi^P_\pi(x)$ and $\phi_\pi^T(x)$ are chirally non-diagonal~\cite{Geshkenbein:1982zs}.
	They are tied by the current identity %(which follows from the relevant Dyson-Schwinger equation)

	\be
	\partial^\nu\big(\overline{u}(0)\sigma_{\mu\nu}\gamma_5d(z)\big)
	=-\partial_\mu\big(\overline{u}(0)i\gamma_5 d(z)\big)+m\,\overline{u}(0)\gamma_\mu\gamma_5d(z)
	\ee
	and share the same couplings.
	The value of the  dimensionful coupling constant $\chi_\pi$ can be fixed by the divergence of the axial-vector  current
	and the PCAC relation

	\begin{eqnarray} \label{eqn_div_of_axial}
		&&(m_u+m_d) \left<0\left|\overline{u}(0)i\gamma^5d(0)\right|\pi^-(p)\right>=\nonumber\\
		&&-(m_u+m_d)\,{\rm Tr}\bigg(i\gamma^5\bigg(\frac{if_\pi}4\gamma^5 \chi_\pi\bigg)\bigg)\,\int_0^1dx\,\phi_{\pi}^P(x)
		=(m_u+m_d)\,f_\pi\chi_\pi
		%\equiv \frac{f_\pi}{\sqrt{2}}m_\pi^2
	\end{eqnarray}
	with $\phi_{\pi}^P(x)$  normalized to 1.  Using the Gell-Mann-Oakes-Renner  relation
	\begin{equation}
		f_\pi^2m_\pi^2=-2(m_u+m_d)\left<\overline{q}q\right>
	\end{equation} 
	with $|\left<\overline q  q\right>|\approx (240\,{\rm MeV})^3$, yield
	\begin{equation}
		\label{CHI}
		\chi_\pi=
		%\frac 1{\sqrt{2}}
		\frac{m_\pi^2}{(m_u+m_d)}
		%\approx  1.2\, {\rm GeV}
		%=-2\sqrt{2}\frac{\left<\overline{\psi}\psi\right>}{f_\pi^2}
	\end{equation}
	The values of the quark masses depend on the renormalization scale $\mu$. Lattice simulations with fine lattices use $\mu\approx 2$ GeV.
	However, for the DAs it is more appropiate to use a softer $\mu\approx 1/\rho$ renormalization with slightly larger current quark masses  giving 
	$\chi_\pi\approx 1.2$ GeV.

	The twist-3 pion DAs asymptote  $\phi_{\pi^-}^P(x)\rightarrow 1$ and $\phi_{\pi^-}^T(x)\rightarrow 6x\bar x$ owing to their conformal collinear spin, with 
	$\phi_{\pi^-}^{T\prime} (x)\rightarrow 6( \bar x- x)$. At large $Q^2$ their contribution is subleading in the pion electromagnetic form factor~\cite{Shuryak:2020ktq}

	\begin{eqnarray} \label{TWIST-2Q}
		\frac{f_\pi^2\chi_\pi^2}{Q^4}  \int  dx_1 dx_2 \, {1\over \bar x_1\bar x_2}
		\bigg[\bigg(\frac 1{\bar x_2}-1\bigg)+(\bar x_2-x_2)\bigg(\frac 1{\bar x_2}+1\bigg)=2\bar x_2\bigg]=2\frac{f_\pi^2\chi_\pi^2}{Q^4}  \int \frac {dx_1}{\bar x_1}\,\,
	\end{eqnarray}

	\section{Twist-3 QDA of the pion and kaon}~\label{QDA-3}
	
	The quasi-pion distribution distribution amplitudes (qPDA) variants of (\ref{MinkDAs}) are 
	
	\begin{subequations}\label{quasiDAs}
		\begin{eqnarray} 
		&&	\tilde{\phi}^A_{\pi^-}(x,P_z)=\frac{i}{f_{\pi}}\int_{-\infty}^{\infty} \frac{dz}{2\pi}e^{i\frac{x-\bar x}{2}P_z z}\langle 0|\bar u (z_-)\gamma^z\gamma^5 [z_-,z_+]d (z_+)|\pi^- (p)\rangle \\
		&& \tilde{\phi}_{\pi^-}^P (x,P_z) = \frac{i P_z}{f_\pi \chi_\pi}\int_{-\infty}^{\infty} \frac{dz}{2\pi}e^{i\frac{x-\bar x}{2}P_z z}\langle 0|\bar u (z_-)\gamma^5 [z_-,z_+]d (z_+)|\pi^- (p) \rangle \\
		&& \tilde{\phi}_{\pi^-}^{T\prime} (x,P_z) = \frac{6P_z}{f_\pi \chi_\pi}\frac{P^{\mu}n'^{\nu}}{P\cdot n'}\int_{-\infty}^{\infty} \frac{dz}{2\pi}e^{i\frac{x-\bar x}{2}P_z z}\langle 0 |\bar u (z_-)\sigma_{\mu\nu}\gamma^5 [z_-,z_+]d (z_+)|\pi^- (p)\rangle
	\end{eqnarray}
	\end{subequations}
	where $z$ is a space-like separation, $z_{\pm}=\pm z/2$, and $n^{\mu}=\{0,0,0,1\}$ is the unit-vector along the linear quark-separation (space-like, $z$-direction here). For the corresponding $K^-$ qPDAs, one would simply replace the $d$-quark with the $s$-quark, and switch to the state $| K^-(P)\rangle $. 
	For finite $P_z$, the qPDAs can be matched with the corresponding light-cone DA counterparts (\ref{MinkDAs}) by an integration kernel, calculable order-by-order in powers of $\frac{\mu}{P_z}$ where $\mu$ represents any other mass-scale present \cite{Ji:2013dva} \cite{Ji:2020ect}. However we will be taking the limit $P_z\rightarrow \infty$, where the matching becomes trivial $\tilde\phi (x,P_z\rightarrow\infty)\rightarrow \phi(x)$. 
	Modulo $x$-independent prefactors, the twist-3 distributions only differ from the twist-2 distribution in their Dirac structure. We write this common factor as:
	
	\begin{equation}\label{qpdafactor}
		\int_{-\infty}^{\infty} \frac{dz}{2\pi}e^{i\frac{x-\bar x}{2}P_z z}\langle 0|\bar u (z_-)\Gamma [z_-,z_+]d (z_+)|\pi^- (p)\rangle
	\end{equation}
	
\begin{figure}
	\centering
	\subfloat{\label{fig_mass:a}\includegraphics[width=.47\textwidth,height=0.32\textwidth]{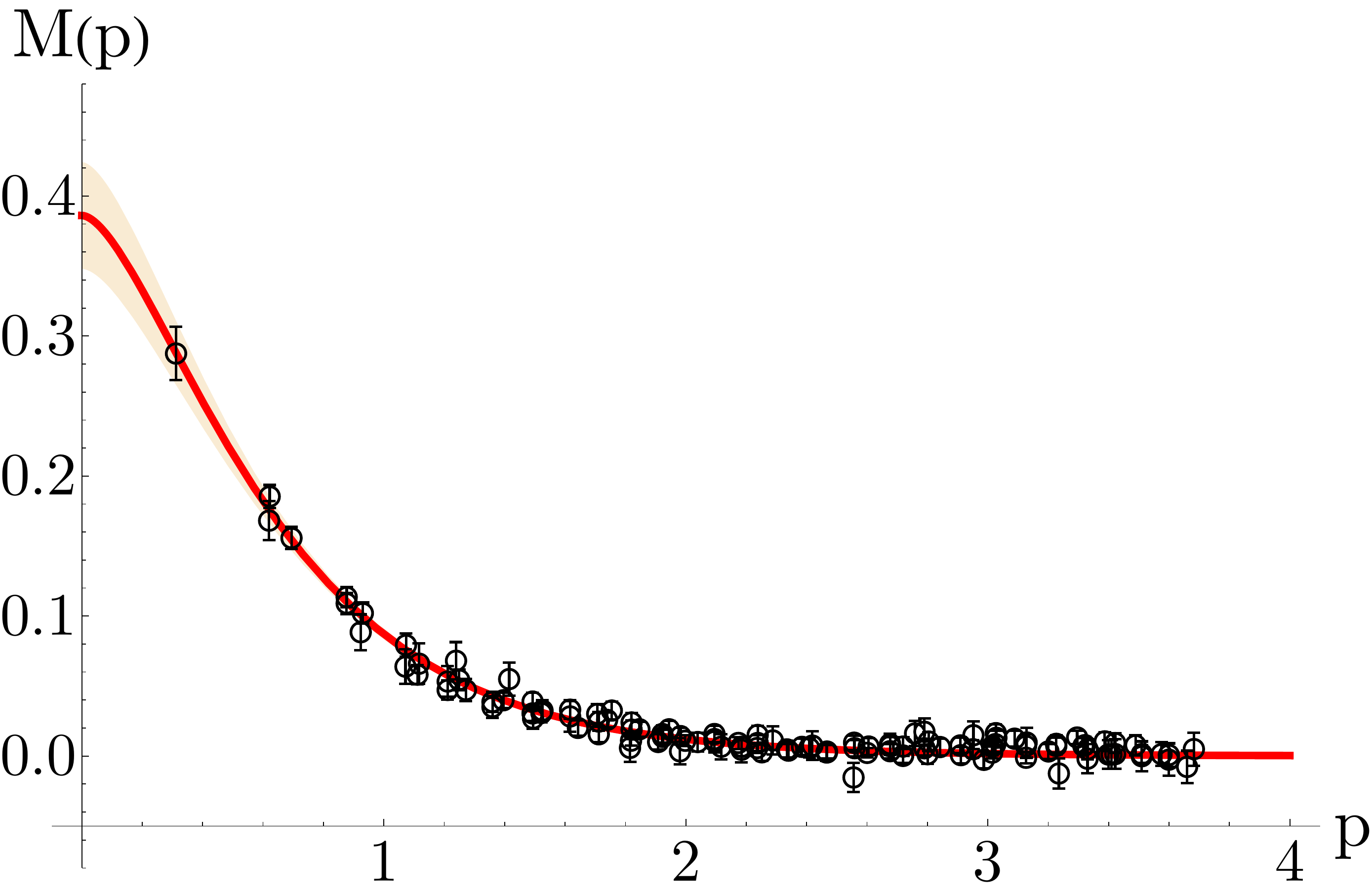}}\hfill
	\subfloat{\label{fig_ConstituentMass:b}\includegraphics[width=.51\textwidth,height=0.32\textwidth]{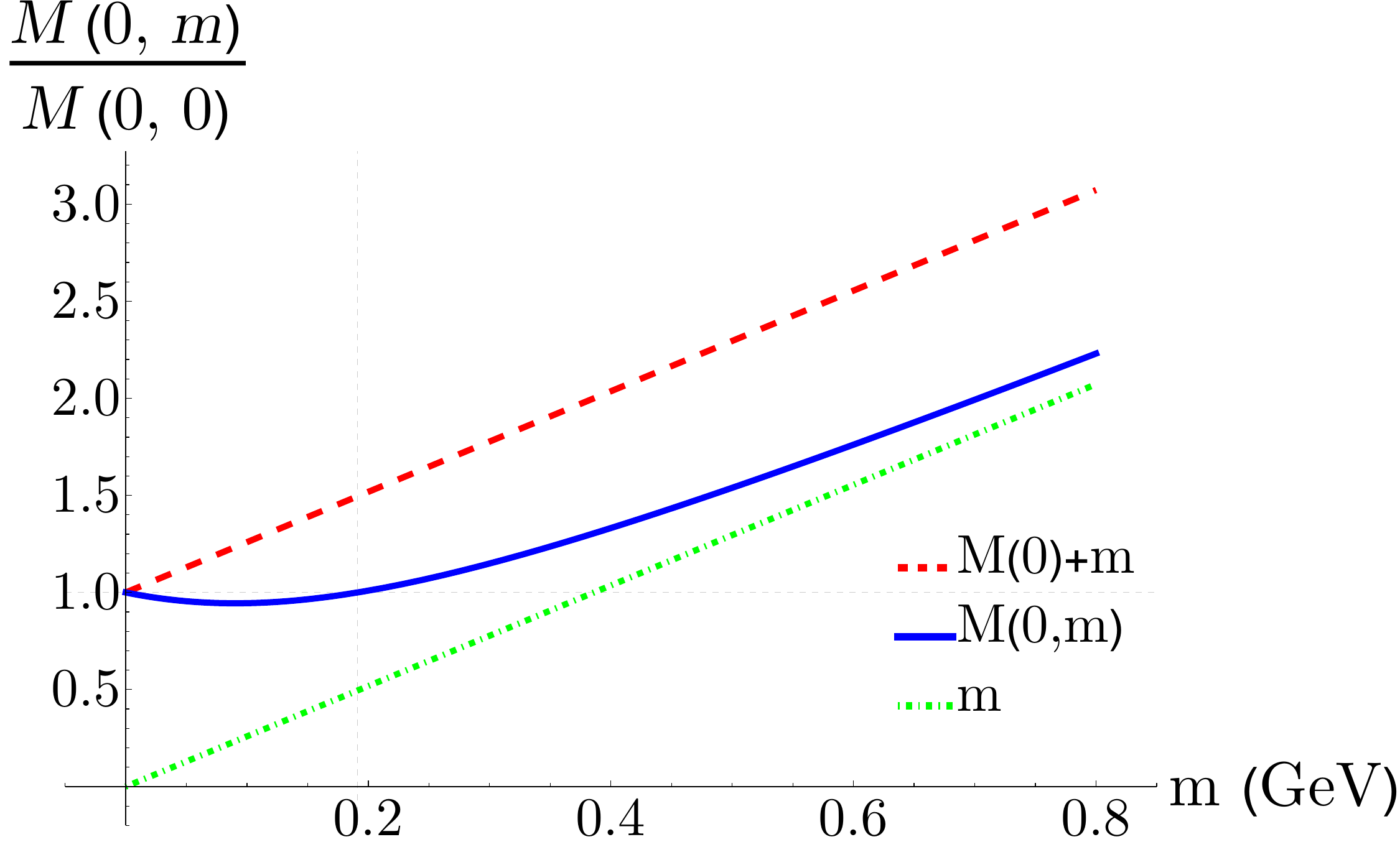}}\hfill
	\caption{ a: Effective quark mass M(p)=M(p,0), both axes GeV; b: Effective quark mass ratio, as a function of current quark mass. See text.}
	\label{fig_MASS}
\end{figure}

	Following the prescription of the present authors' previous paper ~\cite{Kock:2020}, we insert the physical pion source and resum planar diagrams to leading order 
	in the diluteness factor $\alpha\sim \sqrt{\kappa}$ to get

	\begin{equation} \label{qpda1}
		\frac{-P^2}{P_z g_{\pi}}\int \frac{d^4 k}{(2\pi)^4} \delta\left(x-\frac{1}{2}-\frac{k_z}{P_z}\right)
		\textrm{Tr}\left[\Gamma S_1 O_5(P,p_1)S_2\right]
	\end{equation}
where in going from (\ref{qpdafactor}) to (\ref{qpda1}), $S_1$ refers to the $u$-quark and $S_2$ refers to the $d$-quark (or $s$-quark for the negative Kaon). We have subsumed notation for the pion's on-shell condition $\lim P^2\rightarrow m_{\pi}^2$. The trace is over all indices, 
	and $p_{1,2}^{\mu}=k^{\mu}\pm P^{\mu}/2$ is the momentum carried by each quark flavor. The re-summed quark propagator $S_{1,2}\equiv S(p_{1,2},m_{1,2})$ is

	\begin{align}
		S(k,m)&=\bigg(\frac 1{\slashed k -i \sigma(k,m)}\bigg)  \approx \frac{1}{k^2+M^2(0,m)}\left(\slashed k + iM(k,m)\right)
	\end{align}
	where $m$ is the current mass of the individual quark. The effective mass at LO in $\alpha$ is given by $\sigma(k,m)\approx M(k,m) + \mathcal{O}(\alpha^2 )$ 
	
	\begin{align}\label{ConstituentMass}
		\begin{split}
		M(k,m)&=\frac{M(k)}{\left(1+\xi^2\right)^{1/2}+\xi}+m  \\
		M(k)&=M(0)\left(\left|z\left(I_0 K_0-I_1K_1\right)'\right|^2\right)_{z=\frac{k \rho}{2}}\\
		\xi&=\frac{mM(0)\rho^2}{8\pi^2 \kappa} 
		\end{split}
	\end{align}
	with $M(k)\equiv M(k,0)$ throughout.
		 In the above approximation we have dropped the term $\Delta \sigma^2\equiv \sigma^2(k)-\sigma^2(0)$ because it only provides a correction to our final integrals which is subleading in $\alpha$. The benefit of this approximation is that our integrals will have vastly simplified $k^{\mu}$ dependence. 
	In Fig.~\ref{fig_MASS} we show the induced constituent quark mass (\ref{ConstituentMass}) for the parameters of the instanton vacuum. 
	In Fig.~\ref{fig_mass:a} we show  $M(p,0)$ solid-red curve versus $p$
	in GeV units. The spread corresponds to $M(0)=383\pm 39$ MeV and $\rho = 0.313 \pm 0.016$ fm. The open-circles are lattice generated quark masses in 
	Coulomb gauge~\cite{BOWMAN}. In Fig.~\ref{fig_ConstituentMass:b} we show the dependence of the ratio $M(0,m)/M(0,0)$ on the current mass by the solid-blue curve
	for fixed $\xi$, and by the dashed-red curve for $\xi\ll 1$.

	The re-summed pseudoscalar pion vertex is
	
	\begin{equation}
		O_5(P,p_1)\approx \gamma^5\left(1+F_5(P,p_1)\right)+\alpha \bar F_5(P,p_1)+\mathcal{O}(\alpha^2) 
		\end{equation}
	
	\begin{equation}\label{vertex1}
		F_5(P,p_1)\,\overset{P^2\rightarrow m_\pi^2}{\approx} \,\frac{g_{\pi}}{f_\pi} \sqrt{M(p_1)}\frac{1}{P^2+m_{\pi}^2}\sqrt{M(p_2)} 
	\end{equation}
where $g_\pi$ is the pseudoscalar pion-quark-quark coupling. For an explicit calculation of $g_\pi$ in the RIV framework, see section III.C in \cite{Kock:2020}. Expanding to first order in $\alpha$,~[\ref{qpda1}] keeping in mind that $M(k)=\alpha \sigma_0(k)$, the common factor becomes
	
	\begin{align}\label{expansion1}
		\begin{split}	
			\frac{-P^2}{P_z g_{\pi}}\int \frac{d^4 k}{(2\pi)^4}& \,\delta\left(x-\frac{1}{2}-\frac{k_z}{P_z}\right)\\
			\times\Bigg\{ \, &\textrm{Tr}\left[\Gamma\frac{\slashed p_1}{p_1^2+M_1^2}\gamma^5\left(1+F_5(p_1,p_2)\right)\frac{\slashed p_2}{p_2^2+M_2^2}\right]  \\
			& + \textrm{Tr}\left[\Gamma\left(\frac{i\alpha\sigma_0(p_1)}{p_1^2+M_1^2}\right)\gamma^5\left(1+F_5(p_1,p_2)\right)\frac{\slashed p_2}{p_2^2+M_2^2}\right] \\
			& + \textrm{Tr}\left[\Gamma\frac{\slashed p_1}{p_1^2+M_1^2}\gamma^5\left(1+F_5(p_1,p_2)\right)\left(\frac{i\alpha\sigma_0(p_2)}{p_2^2+M_2^2}\right)\right] \\
			& + \textrm{Tr}\left[\Gamma\frac{\slashed p_1}{p_1^2+M_1^2}\alpha \bar F_5 (P,p_1)\frac{\slashed p_2}{p_2^2+M_2^2}\right] \Bigg\} + \mathcal{O}(\alpha^2)
		\end{split}
	\end{align}
	 with $M_{1,2}\equiv M(0,m_{1,2})$ being the effective mass for each quark. 
	 
	 In (\ref{expansion1}) there are four traces: the first is of order $\alpha^0$, the next three are of order $\alpha^1$. If $\Gamma$ contains an odd number of $\gamma$'s (e.g. $\Gamma=\gamma^{z}\gamma^5$), the first trace term at order $\alpha^0$ has vanishing Dirac trace, whereas the remaining three do not have vanishing Dirac trace. This is the case when calculating the twist-2 qPDA. 
	However if $\Gamma$ contains an even number of $\gamma$'s (e.g. $\Gamma=\gamma^5,\,\sigma_{\mu\nu}\gamma^5$), then the first term has nonvanishing Dirac trace. 
	At next to leading order (NLO) in $\alpha$, the second and third terms vanish. The fourth term involving $\bar F_5$ does not have vanishing Dirac trace, and requires special attention. This is the case with the twist-3 distributions, which we are considering here. In our previous paper \cite{Kock:2020} we showed that this term vanishes for the axial-vector twist-2 DA, $\Gamma=\gamma^z\gamma^5$. We now make explicit the leading contributions in $\alpha\sim \sqrt{\kappa}$ 
	to the twist-3 DAs, $\phi^P_0(x)$ and $\phi^T_0(x)$. We also recap the similar expression for the twist-2 DA, $\phi^A_0(x)$.

	\subsection{Pseudoscalar, $\Gamma=\gamma^5$}
	
	The calculation of $\tilde \phi^P_0(x,)$ begins by reinstating the appropriate prefactor in (\ref{expansion1}). Out of the first three zero-mode contributions in (\ref{expansion1}), only the first has non-vanishing Dirac trace
	
	\begin{align}\label{qpda01}
		\tilde \phi^P_0(x,P_z)&=\frac{iP_z}{f_\pi\chi_\pi}\frac{-P^2}{P_z g_\pi} \int \frac{d^4 k}{(2\pi)^4} \,\delta\left(x-\frac{1}{2}-\frac{k_z}{P_z}\right)
		\textrm{Tr}\left[\gamma^5\frac{\slashed p_1}{p_1^2+M_1^2}\gamma^5\left(1+F_5(p_1,p_2)\right)\frac{\slashed p_2}{p_2^2+M_2^2}\right] \notag \\
		& = \frac{4iN_c}{f_\pi^2\chi_{\pi}}\int \frac{d^2k_\perp dk_4 dk_z}{(2\pi)^4}\delta\left(x-\frac{1}{2}-\frac{k_z}{P_z}\right) \left(M(p_1)M(p_2)\right)^{1/2}\frac{p_1\cdot p_2}{(p_1^2+M_1^2)(p_2^2+M_2^2)} \notag \\
		&=  \frac{4iN_cP_z}{f_\pi^2\chi_{\pi}}\int \frac{d^2k_\perp dk_4}{(2\pi)^4} \left(M(p_1)M(p_2)\right)^{1/2}\frac{p_1\cdot p_2}{(p_1^2+M_1^2)(p_2^2+M_2^2)}
	\end{align}
where the delta-function has set $P_{1,z}=x P_z$ and $P_{2,z}=\bar x P_z$.  Since
	
	\begin{equation}
		p_1-p_2=P\implies p_1\cdot p_2 = \frac{1}{2}\left[p_1^2+p_2^2-m_{\pi}^2\right]
	\end{equation}
	the last term in (\ref{qpda01}) can be recast in the form
	
	\begin{equation}
	\label{PHIP}
		\tilde \phi^P_0(x,P_z) = \frac{2iN_cP_z}{f_\pi^2\chi_{\pi}}\int \frac{d^2k_\perp dk_4}{(2\pi)^4} \left(M(p_1)M(p_2)\right)^{1/2}\frac{p_1^2+p_2^2-m_\pi^2}{(p_1^2+M_1^2)(p_2^2+M_2^2)}
	\end{equation}
Notice that for the $k_4$-integration, the pole and branch point structure is exactly the same as in the twist-2 case. Therefore we evaluate in the same way, by Wick-rotating and shifting in $k_4$, into the Minkowski domain as we detail in appendix~\hyperref[appendixA]{A}.
		Taking the final limit $P_z\rightarrow \infty$, including the finite pion mass~\cite{Kock:2020},
	and simplifying the pre-factor  using $\chi_{\pi}=-2\langle\bar \psi \psi \rangle/f_{\pi}^2$, the final result for the leading-order pseudoscalar DA is

	\begin{equation}\label{qpda3}
		\tilde \phi^P_0(x)=i\frac{N_c}{2\langle\bar \psi \psi \rangle} \frac{\theta(x\bar x)}{x\bar x}\int\frac{d^2k_{\perp}}{(2\pi)^3}\, M\left( \frac{k_{\perp}}{\lambda \sqrt{x \bar x }}\right) \frac{k_\perp^2+\bar x^2M_1^2+x^2M_2^2}{k_\perp^2+\bar x M_1^2+xM_2^2-x \bar x m_{\pi}^2}
	\end{equation}

\begin{figure}[ht]
		\centering
		\subfloat{\label{fig_QPDA_ChangingRho:a}\includegraphics[width=.49\textwidth,height=50mm]{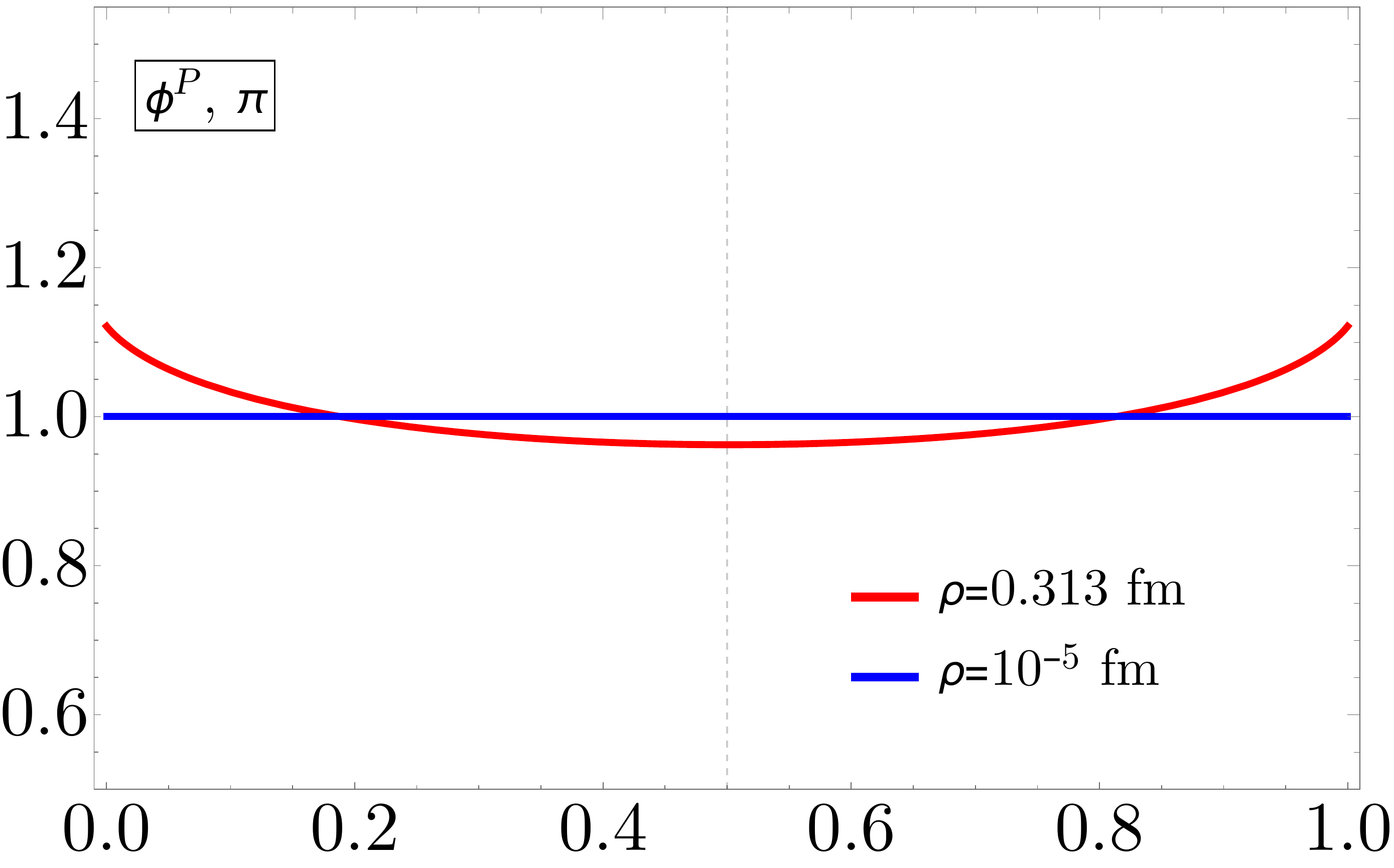}}\hfill
		\subfloat{\label{fig_QPDA_ChangingRho:b}\includegraphics[width=.49\textwidth,height=50mm]{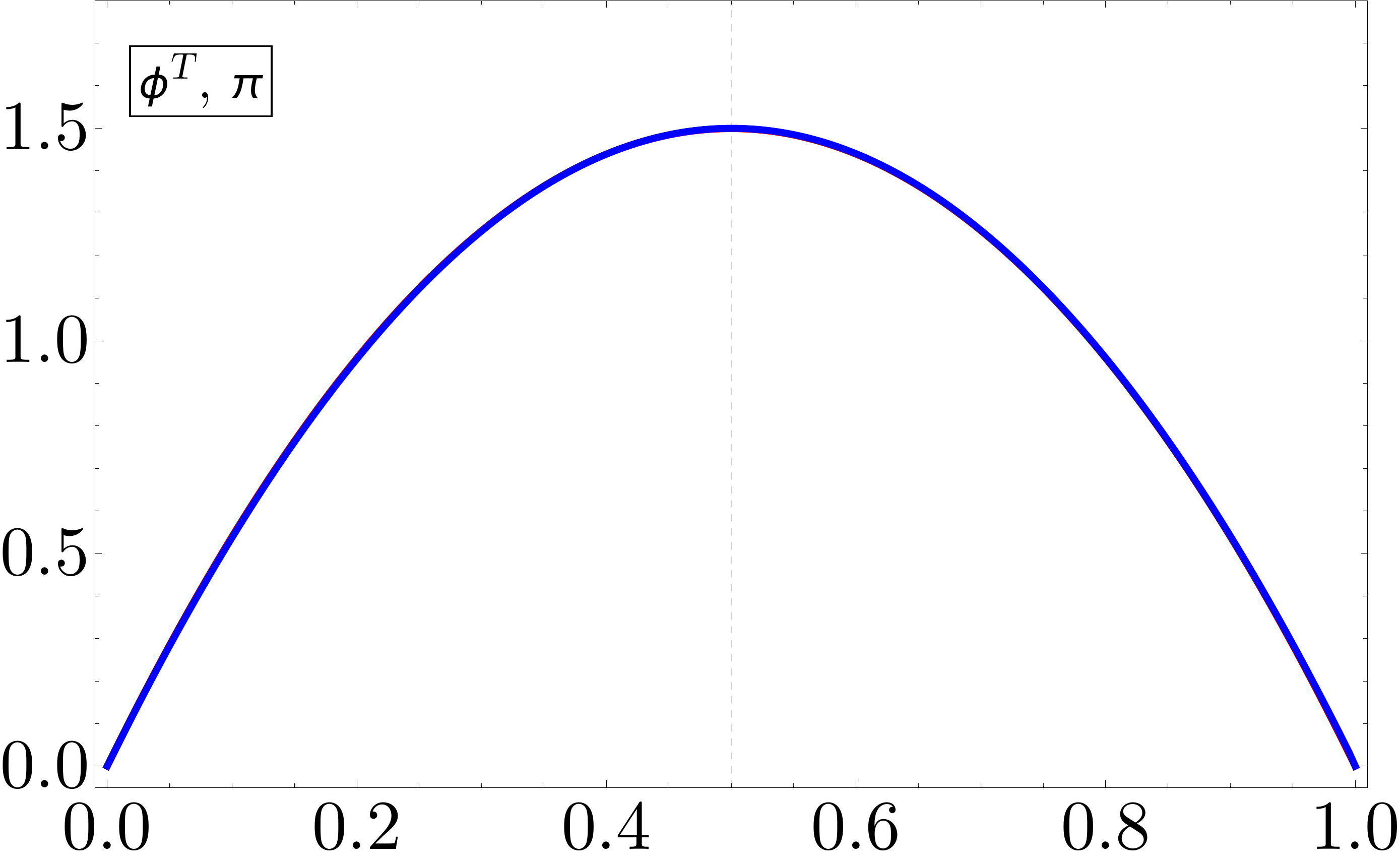}}\hfill
		\caption{Pseudoscalar and pseudotensor pion DAs at different instanton sizes or resolution. The $\rho\approx 0$ curves (solid blue) are indiscernible from the asymptotic forms. For the pseudotensor DA $\phi^T$, both for phenomenological and small $\rho$, the curves are indistinguishable from the asymptotic form $6x\bar x$.}
		\label{fig_QPDA_ChangingRho}
	\end{figure}

	The requisite $x\leftrightarrow \bar x$, $m_1\leftrightarrow m_2$ symmetry is manifest. For plotting we use phenomenological values $M(0)= 383\,\textrm{MeV}$, $|\langle\bar\psi \psi\rangle |=(240\,\textrm{MeV})^3$, and $\rho=0.313\,\textrm{fm}$. The normalizing values of $\lambda$ are provided in table \ref{lambda_table} in appendix \hyperref[appendixB]{B}. Unevolved plots of (\ref{qpda3}) are shown in Fig. \ref{fig_QPDA_ChangingRho} for both the limit of a small instanton size $\rho\rightarrow 0$ and typical instanton size $\rho=0.313\,\textrm{fm}$. For the pion, there is no perceptible difference in (\ref{qpda3}) between using physical masses ($m_{\pi}=140\,\textrm{MeV}$, $m_u=3\,\textrm{MeV}$, $m_d=7\,\textrm{MeV}$) and the chiral counterparts ($m_{\pi,u,d}=0$). For the kaon we use $m_{K}=494\,\textrm{MeV}$, $m_u=3\,\textrm{MeV}$, $m_s=136\,\textrm{MeV}$. The small-size instanton limit ($\rho\rightarrow 0$) corresponds to very high resolution and is commensurate with the QCD asymptotic result as expected.

	\subsection{Pseudotensor, $\Gamma=\sigma_{\mu\nu}\gamma^5$}

	The only difference with the pseudoscalar case will be that instead of having a factor $i\textrm{Tr}\left[\slashed{p_1}\slashed{p_2}\right]$, we will have
	
	\begin{align}
		6 \frac{P_{\mu}n_{\nu}}{P\cdot n}\textrm{Tr}\left[\sigma^{\mu\nu}\slashed{p_1}\slashed{p_2}\right]
		&= -24 i \frac{P_{\mu}n_{\nu}}{P\cdot n} \left(p_1^\mu p_2^\nu - p_2^\mu p_1^\nu \right) \notag \\
		&=\frac{-24 i}{P\cdot n} \left[(P\cdot p_1)(n\cdot p_2)-(n\cdot p_1)(P\cdot p_2)\right]
	\end{align}
where $n^{\mu}$ is the tangent vector to the spacelike quark separation line in~(\ref{quasiDAs}), $n^{\mu}=\hat z^{\mu}$. Making this replacement in~(\ref{qpda01}), we get
	
	\begin{equation}
		\tilde \phi^{T\prime}_0(x,P_z) =  \frac{-24 i N_c }{f_\pi^2\chi_{\pi}}\int \frac{d^2k_\perp dk_4}{(2\pi)^4} \left(M(p_1)M(p_2)\right)^{1/2}\frac{(P\cdot p_1)(n\cdot p_2)-(n\cdot p_1)(P\cdot p_2)}{\left(p_1^2+M_1^2\right)\left(p_2^2+M_2^2\right)}
	\end{equation}
As before, we Wick-rotate and shift the $k_4$ integration, $k_4\rightarrow i(k_4+(x-\frac{1}{2}E))$, leaving us with 
	
	\begin{equation}
		\tilde \phi^{T\prime}_0(x,P_z) =  \frac{24 N_c}{f_\pi^2\chi_{\pi}}\int \frac{d^2k_\perp dk_4}{(2\pi)^4} \left(M(y_1)M(y_2)\right)^{1/2}\frac{(P\cdot y_1)(n\cdot y_2)-(n\cdot y_1)(P\cdot y_2)}{\left(y_1^2+M_1^2\right)\left(y_2^2+M_2^2\right)}
	\end{equation}
	To evaluate the integrand we use the same kinematics as in~(\ref{kinematics}). To perform the $k_4$ integral, we follow the same procedure as in the pseudoscalar case - use the modified effective mass, then integrate the remaining rational function using Cauchy's residue theorem. The final result for the integrated $\phi^T(x)$ is
	
	\begin{align}\label{qpda5}
		\tilde \phi^{T}_0(x)&=\frac{-3 i N_c}{\langle \bar \psi \psi \rangle}\theta (x\bar x) \int_0^x dv \,\int \frac{d^2 k_{\perp}}{(2\pi)^3} M\left(\frac{k_{\perp}}{\lambda\sqrt{v\bar v}}\right)\frac{1}{v\bar v}\frac{(\bar v - v)k_{\perp}^2+\bar{v}^2 M_1^2 - v^2 M_2^2}{k_{\perp}^2-v\bar v m_{\pi}^2+\bar v M_1^2 + v M_2^2} \notag \\
		&
	\end{align}
Again, the requisite $x\leftrightarrow \bar x$, $m_1\leftrightarrow m_2$ symmetry is manifest (the integrand is odd under this transformation). Unevolved plots of (\ref{qpda5}) are shown in Fig. \ref{fig_QPDA_ChangingRho} for phenomenological and limiting values of $\rho$. In the case that $\rho\rightarrow 0$ (high resolution), the pseudotensor DA approaches its asymptotic form.  Once again, the normalizing values of $\lambda$ are given in table \ref{lambda_table} in appendix \hyperref[appendixB]{B}.

\begin{figure}[ht]
		\centering
		\subfloat{\label{fig_QPDA_ChangingRho_Twist2:a}\includegraphics[width=.49\textwidth,height=50mm]{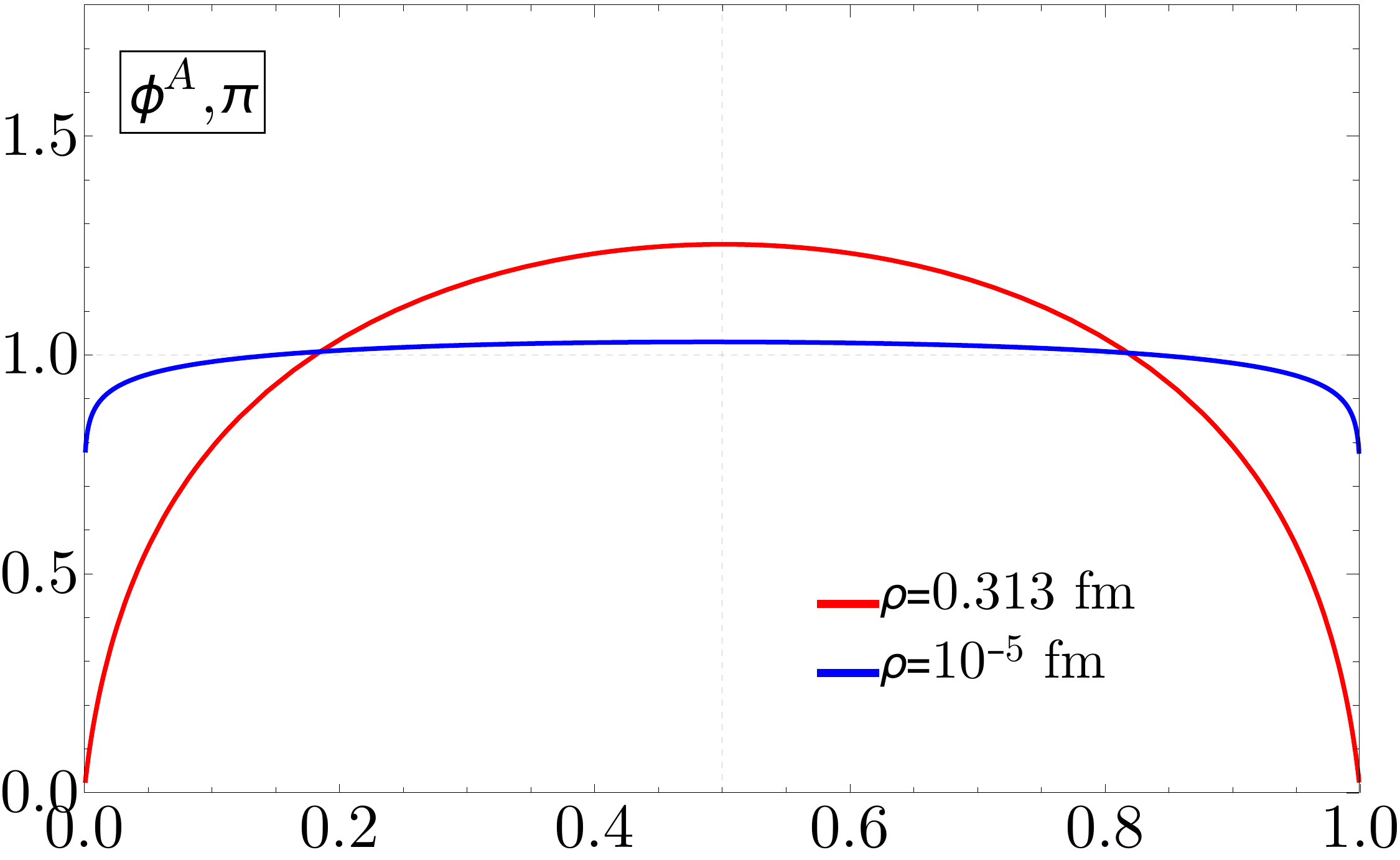}}\hfill
		\subfloat{\label{fig_QPDA_ChangingRho_Twist2:b}\includegraphics[width=.49\textwidth,height=50mm]{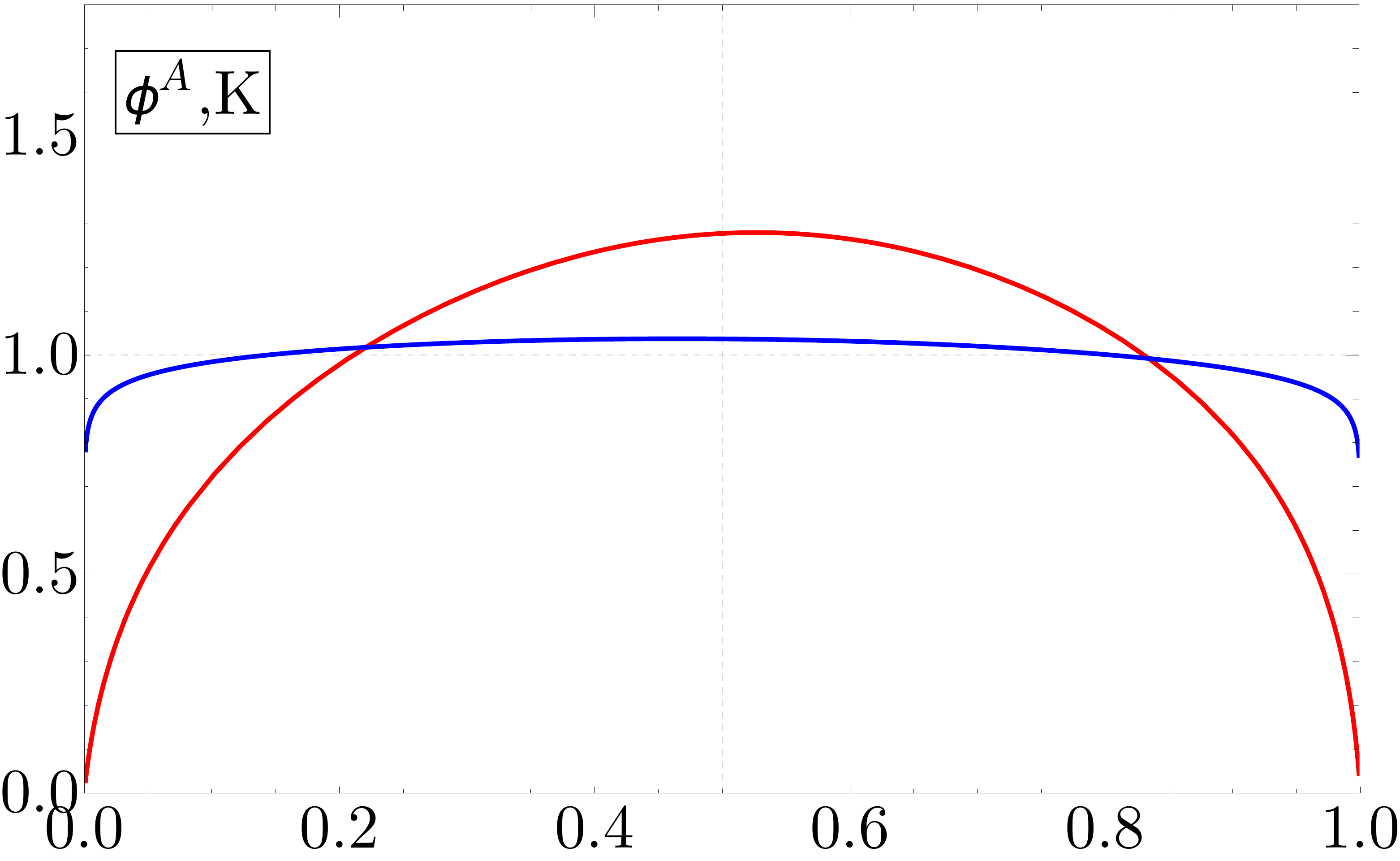}}\hfill
		\caption{Twist-2 (axial-vector) pion and kaon DAs at different instanton sizes or resolution. The normalizing values of $\lambda$ for the $\rho=10^{-5}\,\textrm{fm}$ curves are as follows: $\lambda_{\pi}=7.51$, $\lambda_K=9.7$ (for the phenomenological curves see Appendix \ref{appendixB}). For both the pion and kaon, the $\rho\approx 0$ curves (solid blue) tend toward a normalized step-function $\theta(x\bar x )$.}
		\label{fig_QPDA_ChangingRho_Twist2}
	\end{figure}

	\subsection{Axial-vector, Twist-2, $\Gamma=\gamma^z\gamma^5$}
	
	Here we generalize a key expression from our previous paper \cite{Kock:2020}: the twist-2 DA at leading order in $\alpha$, now including finite current quark masses $m_{1,2}$ for the pseudoscalar meson $P$. It is given by
	
	\begin{equation}\label{twist2}
		\phi^A_P (x) \approx \frac{2N_c}{f_P^2}\theta (x\bar x ) \int \frac{d^2k_\perp}{(2\pi)^3} M\left(\tilde k_\perp\right)\,\frac{\bar x M \left(\tilde k_\perp,m_1\right)+x M \left(\tilde k_\perp,m_2\right)}{k_\perp^2- 	x \bar x m_P^2 + \bar x M_1^2 + x M_2^2}
	\end{equation}
	with $\tilde k_\perp = {k_\perp}/{\lambda \sqrt{ x \bar x}}$.
	We use $f_\pi = 130\,\textrm{MeV}$ for massive pions and $f_K = 155\,\textrm{MeV}$ for massive kaons. Unevolved plots of (\ref{twist2}) are shown in Fig. \ref{fig_QPDA_ChangingRho_Twist2} for phenomenological and limiting values of $\rho$. For $\rho \rightarrow 0$ the curves tend towards a normalized step-function, rather than towards the asymptotic distribution $6x\bar x$. This type of curve has been noted for chiral quark models with point interactions \cite{Broniowski:2017wbr} \cite{Jia:2019ary}, and some bound-state resummations \cite{Ding2020}.

	\begin{figure}
		\centering
		\subfloat{\label{fig_PDAEVL:a}\includegraphics[width=.49\textwidth,height=0.32\textwidth]{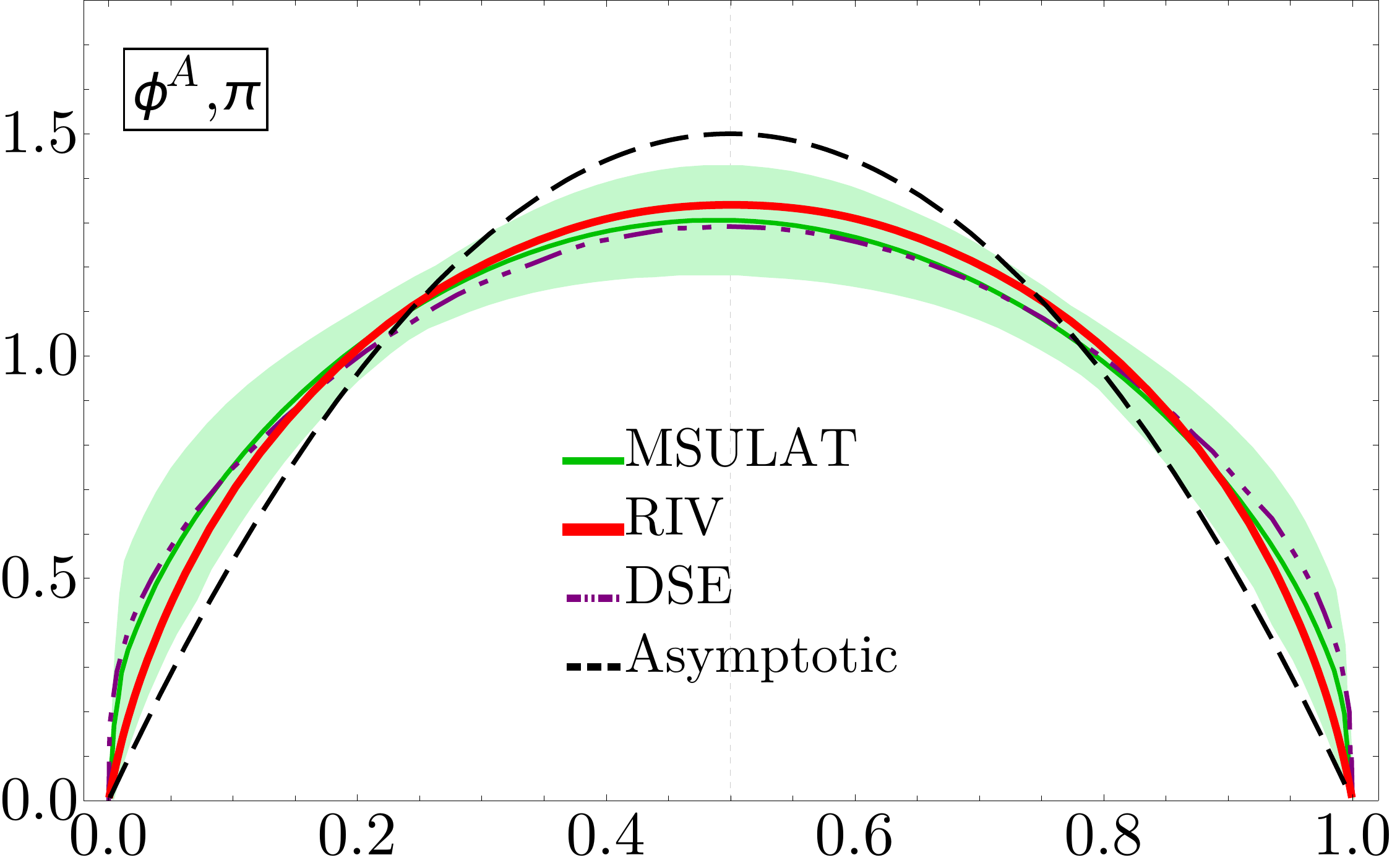}}\hfill
		\subfloat{\label{fig_PDAEVL:b}\includegraphics[width=.49\textwidth,height=0.32\textwidth]{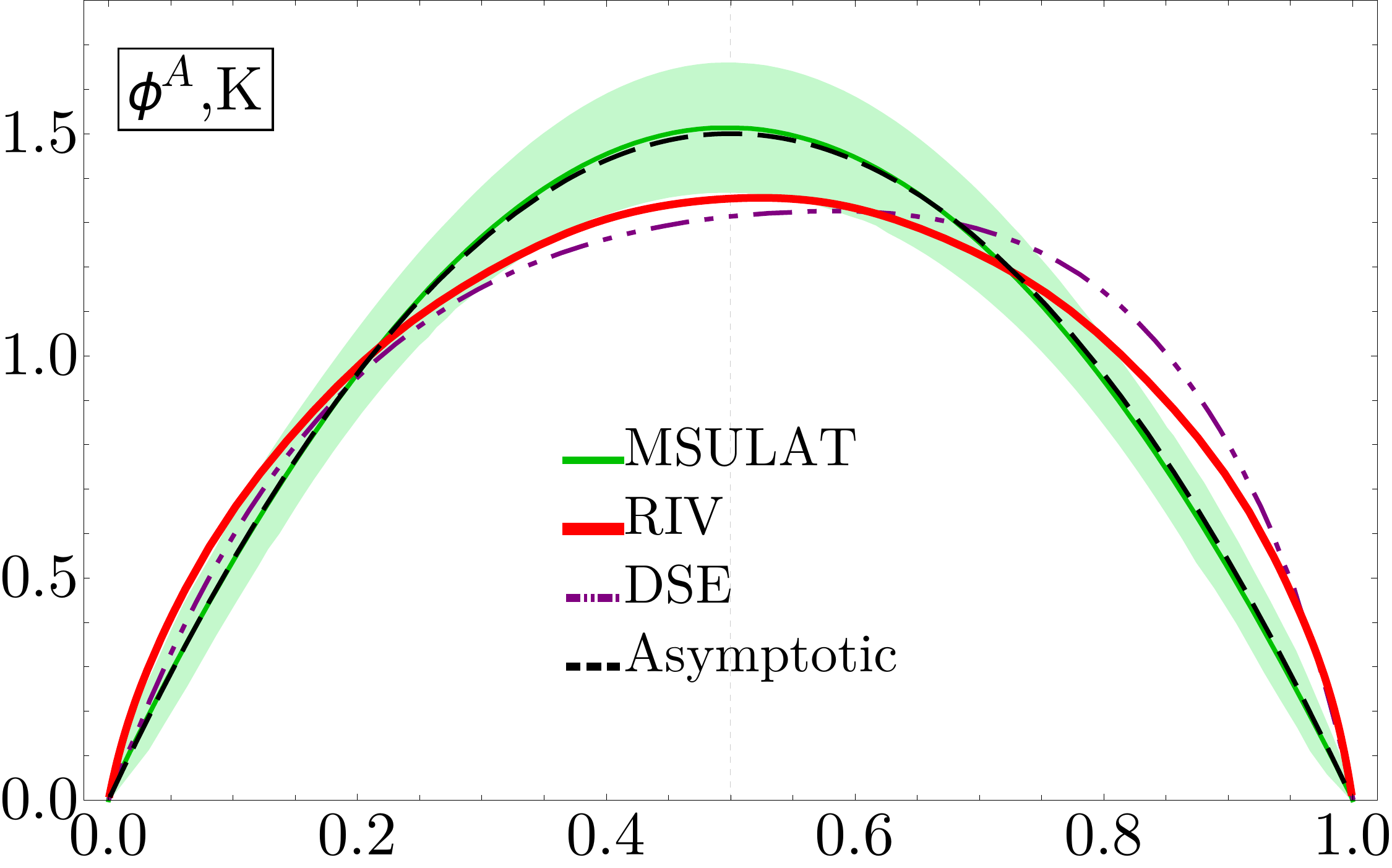}}\hfill
		\subfloat{\label{fig_PDAEVL:c}\includegraphics[width=.49\textwidth,height=0.32\textwidth]{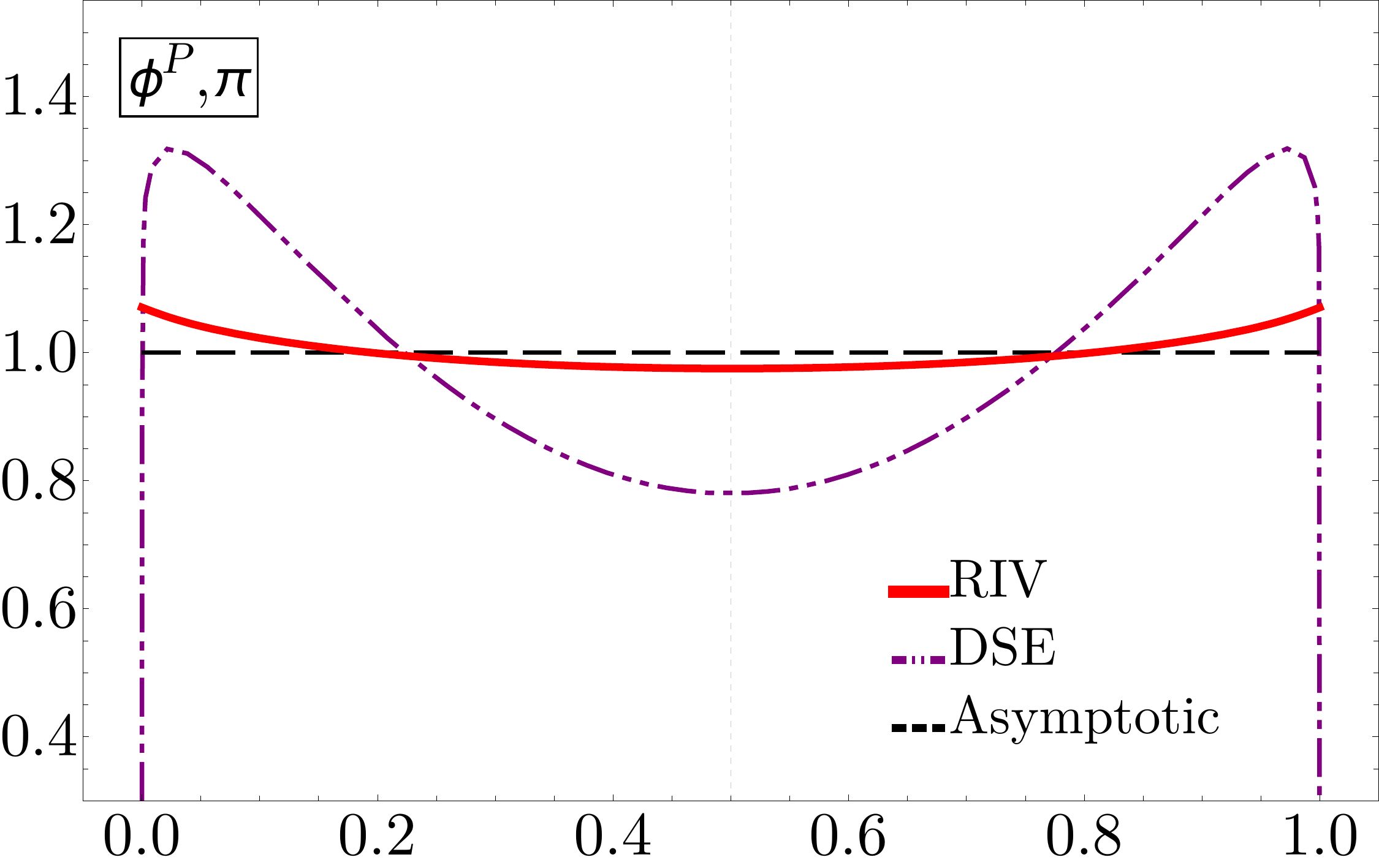}}\hfill
		\subfloat{\label{fig_PDAEVL:d}\includegraphics[width=.49\textwidth,height=0.32\textwidth]{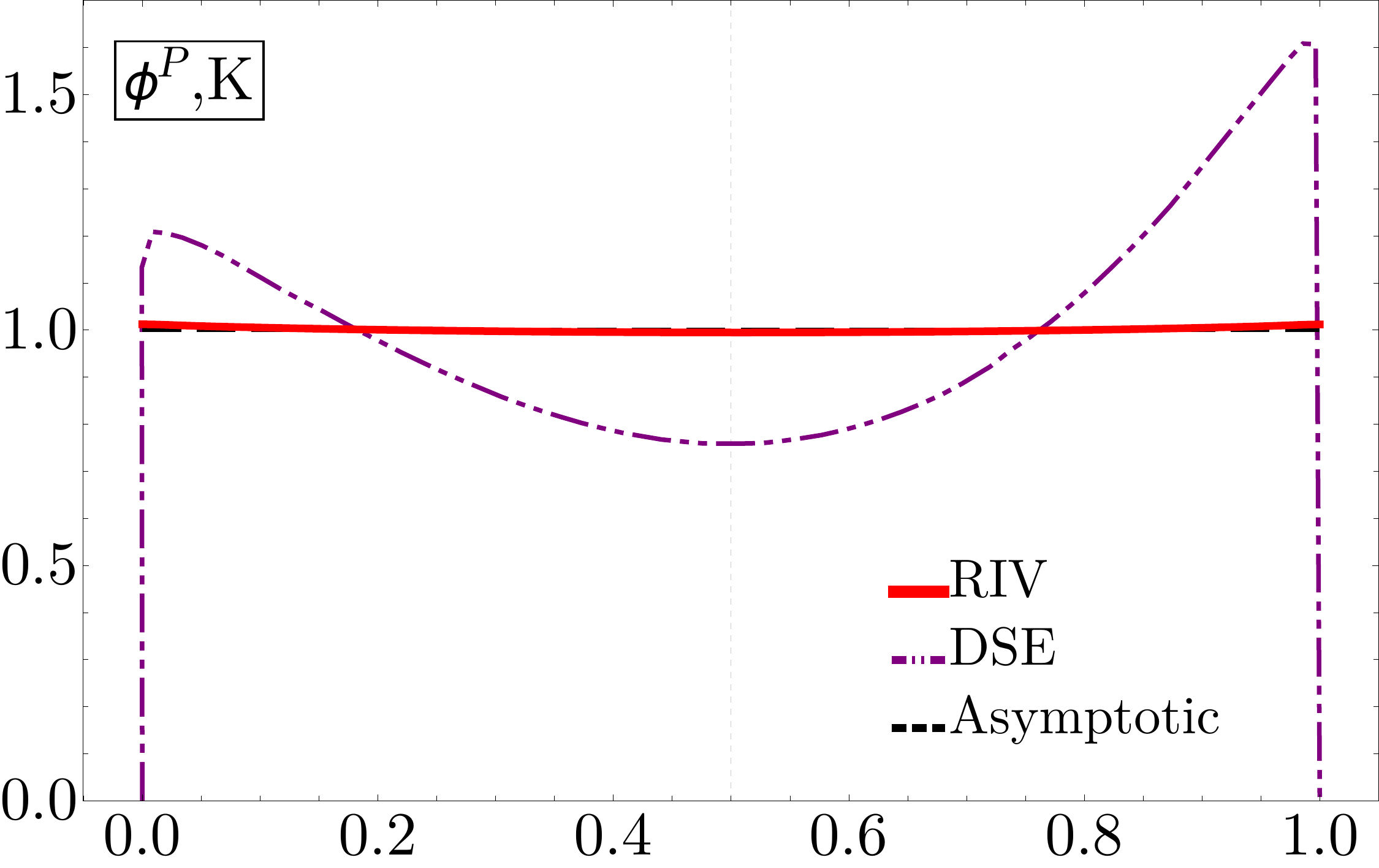}}\hfill
		\subfloat{\label{fig_PDAEVL:e}\includegraphics[width=.49\textwidth,height=0.32\textwidth]{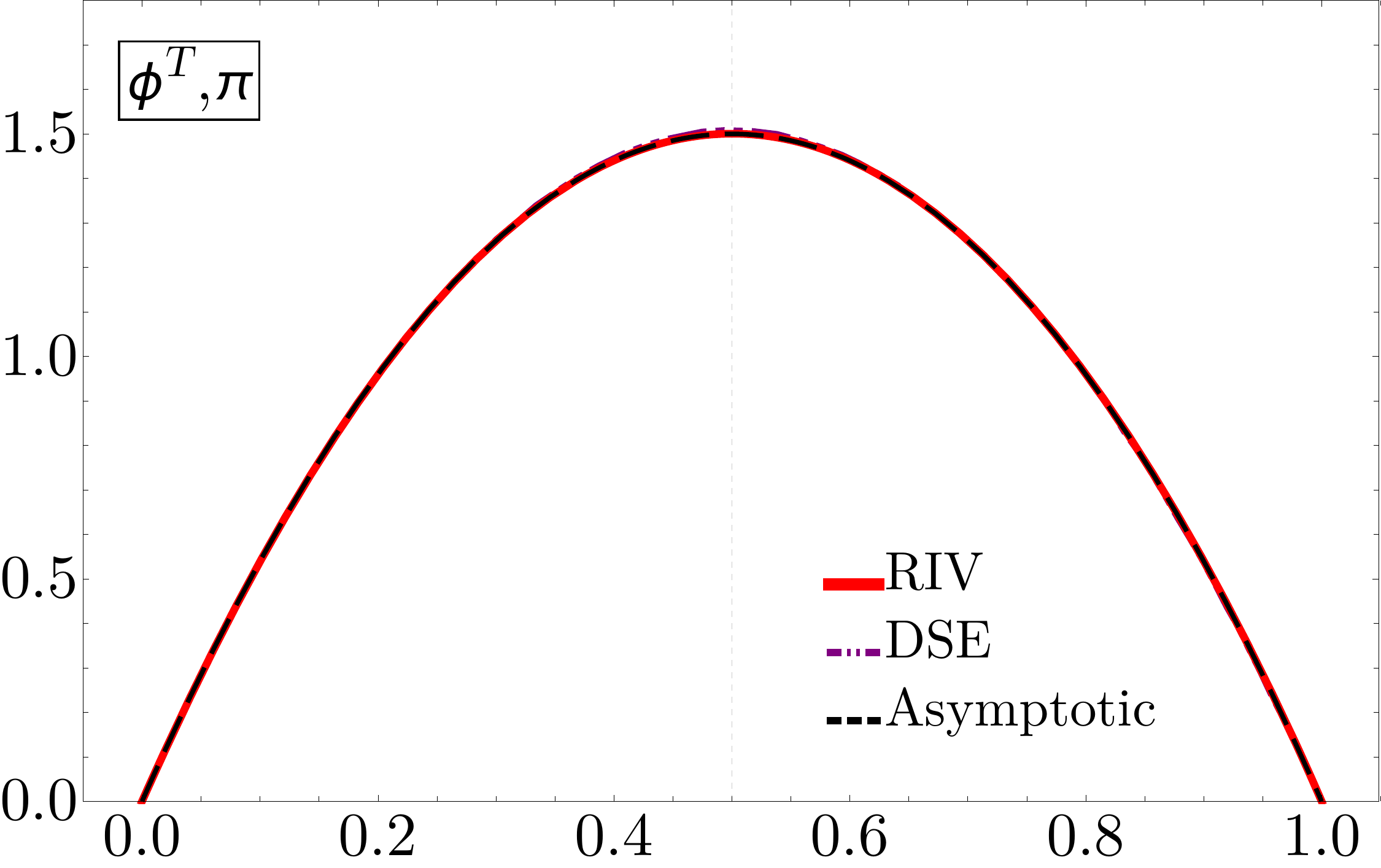}}\hfill
		\subfloat{\label{fig_PDAEVL:f}\includegraphics[width=.49\textwidth,height=0.32\textwidth]{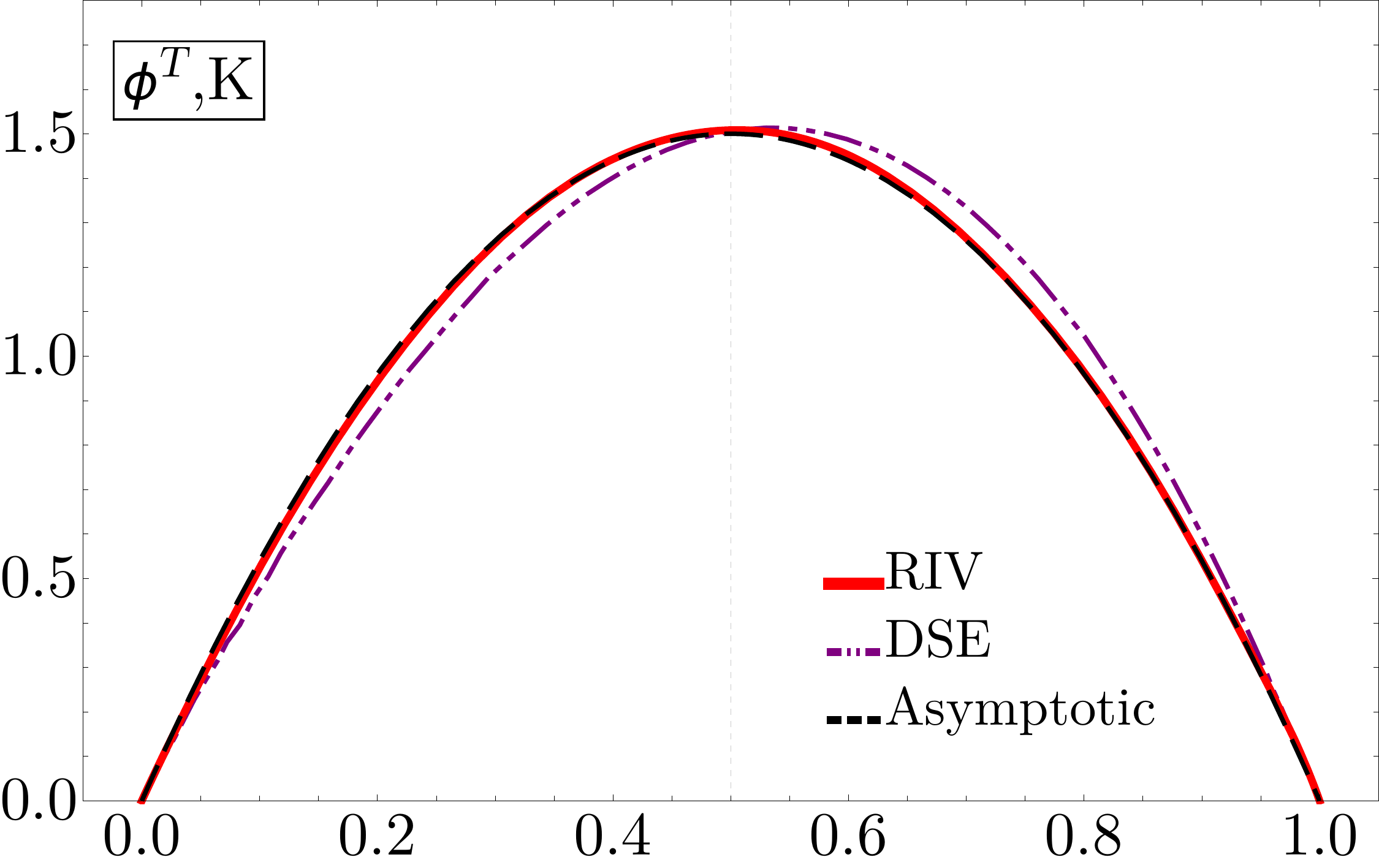}}\hfill
		\caption{ERBL evolution of DAs, compared with data. All curves are given at $Q=2\,\textrm{GeV}$, except for LFCQM \cite{deMelo2016} (blue-dotted) which is renormalized in a different scheme. RIV (red-solid-thick) represents our current work, the random instanton vacuum approach, given by equations (\ref{qpda3}), (\ref{qpda5}), and (\ref{twist2}). MSULAT curves (green-solid-bands) are recent lattice computations utilizing the LaMET framework \cite{Zhang2020}. DSE curves (purple-dash-dot-dot) are from \cite{Shi2015} which utilized Dyson-Schwinger equations with Bethe-Salpeter amplitudes.}
		\label{fig_PDAEVL}
	\end{figure}

	\section{QCD Evolution}~\label{evolution}
	
	The two-particle twist 2 \& 3 DAs in the random instanton vacuum (RIV) are defined 
	at a low renormalization scale set by the typical  inverse instanton size $Q_0 = 1/\rho = 631$ GeV. Assuming factorization, their forms at higher renormalization scales follow from QCD  evolution equations

	\begin{subequations}
		\begin{align}
		\phi^A(x,Q) &=6x\bar x \sum_{n=0}^{\infty} a_n(Q_0) \left(\frac{\alpha_s(Q^2)}{\alpha_s(Q^2_0)}\right)^{\gamma_n^A/\beta_0}	 C_n^{3/2}(x-\bar x) \\
		\phi^P(x,Q) &= 1+\sum_{n=1}^{\infty} b_n(Q_0) \left(\frac{\alpha_s(Q^2)}{\alpha_s(Q^2_0)}\right)^{\gamma_n^P/\beta_0}	 C_n^{1/2}(x-\bar x) \\
		\phi^T(x,Q) &= 6x\bar x\left[1+ \sum_{n=1}^{\infty} c_n(Q_0) \left(\frac{\alpha_s(Q^2)}{\alpha_s(Q^2_0)}\right)^{\gamma_n^T/\beta_0}	 C_n^{3/2}(x-\bar x) \right]
		\end{align}
	\end{subequations}
with the anomalous dimensions $\gamma_n^{A,P,T}$ given by \cite{Shifman1981FormFO}

	\begin{subequations}
		\begin{align}
		\gamma_n^A &= C_F\left[-3+4\sum_{j=1}^{n+1}\frac{1}{j} - \frac{2}{(n+1)(n+2)}\right] \\
		\gamma_n^P &= C_F\left[-3+4\sum_{j=1}^{n+1}\frac{1}{j} - \frac{8}{(n+1)(n+2)}\right] \\
		\gamma_n^T &= C_F\left[-3+4\sum_{j=1}^{n+1}\frac{1}{j}\right]
		\end{align}
	\end{subequations}
Here $C_n^m(z)$ are Gegenbauer polynomials, $C_F=(N_c^2-1)/2N_c$ is the quadratic-Casimir in the fundamental representation, $\alpha_s(Q^2)=4\pi/\left( \beta_0\ln (Q^2/\Lambda^2)\right)$ is the one-loop running QCD coupling, $\beta_0=\frac{11}{3}N_c - \frac{2}{3} N_f$, and $\Lambda=250\,\textrm{MeV}$. 
	%Our initial scale is given by the instanton size, $Q_0=1/\rho = 631\,\textrm{MeV}$. 
	One can easily verify that the normalizations are preserved under QCD  evolution, as they should be. Owing to the orthogonality of the Gegenbauer polynomials, the initial coefficients are given by
	
	\begin{subequations}
		\begin{align}
		a_n(Q_0) &= \frac{2(2n+3)}{3(n+1)(n+2)}\int_0^1 dy \, C_n^{3/2}(y-\bar y) \phi^A(y,Q_0)\\
		b_n(Q_0) &= (2n+1)\int_0^1 dy\, C_n^{1/2}(y-\bar y) \phi^P(y,Q_0)\\
		c_n(Q_0) &= \frac{2(2n+3)}{3(n+1)(n+2)}\int_0^1 dy \, C_n^{3/2}(y-\bar y) \phi^T(y,Q_0)
		\end{align}
	\end{subequations}

	The twist-2 \& twist-3 DAs, evolved to $Q=2\,\textrm{GeV}$, are shown in fig.\ref{fig_PDAEVL}. All curves are shown at the same renormalization scale.
	Our twist-2 pion DA shows a shape slightly broader than the asymptotic form. This shape has been seen in recent lattice calculations \cite{Zhang:2017bzy}\cite{Zhang2020}. In some light-front constituent quark models, a shape slightly narrower than the asymptotic form is seen - we do not display this curve in fig.\ref{fig_PDAEVL:a} because of a mismatch in renormalization schemes \cite{deMelo2016}. The empirical pion twist-2 DA extracted from dijet data by the E791 collaboration is in agreement with all curves shown fig.\ref{fig_PDAEVL:a}, though the precise shape is obscured by uncertainties \cite{E791}. The same broad shape is seen in our twist-2 kaon DA, fig.\ref{fig_PDAEVL:b}, although we see a smaller asymmetry than other phenomenological approaches.

	The general behavior of all our DAs show agreement with those denoted DSE \cite{Shi2015}, except that our curves are closer to the respective asymptotic forms. This is most notable in the pseudoscalar DAs, where our curves are remarkably closer to the asymptotic form. Although our kaon's pseudoscalar DA seems to lack asymmetry, especially compared to the pion's pseudoscalar DA, this is only because the overall scale of all its Gegenbauer moments are smaller, thereby making its difference from the asymptotic form indiscernible. We can see this with a comparison of the first two non-trivial Gegenbauer moments - the ratio $b_1/b_2$ is nearly an order of magnitude larger for the kaon compared to the pion.
	
	\begin{subequations}
		\begin{align}
		b_1^{\pi}(Q_0)&= -8.4\times 10^{-5} & b_2^{\pi}(Q_0)&= 1.74\times 10^{-2} & \left| \frac{b_1^{\pi}(Q_0)}{b_2^{\pi}(Q_0)}\right| &= 4.82\times 10^{-3}\\
		b_1^{K}(Q_0)&= -1.41\times 10^{-4} & b_2^{K}(Q_0)&= 3.46 \times 10^{-3} & \left| \frac{b_1^{K}(Q_0)}{b_2^{K}(Q_0)}\right| &= 4.07\times 10^{-2} 
		\end{align} 
	\end{subequations}

	\section{Conclusions}
	\label{sec_conclusions}

	Cooled lattice gauge configurations display strongly inhomogeneous instanton and anti-instanton configurations. The  dilute QCD instanton vacuum 
	in its simplified RIV form  capture the essentials physics of these tunneling configurations at low resolution. Each tunneling traps a zero mode of a given
	chirality for each flavor, breaking dynamically chiral symmetry. The disordering of these zero modes leads to a multitude of multiquark condensates
	and a running constituent quark mass~\cite{Schafer:1996wv,Zahed:2021fxk}.
	
	In the RIV the pion quasi-DA is a state made of zero-modes and non-zero-modes that interact collectively. While still complex, this state can be organized
	in terms of the RIV diluteness factor $\alpha\sim \sqrt{\kappa}$. In leading order in $\alpha$, the twist-2 contribution to the pion quasi-DA involves both the
	zero-modes and non-zero-modes as we have analyzed in details in~\cite{Kock:2020} and smoothly yields the pion DA in the large momentum limit. 
	quasi-DA are dominated solely by the zero-modes owing to their pseudo-scalar and pseudotensor content. They also yield smoothly the pion DA in the large
	momentum limit. In all cases the DA follows from  the large momentum limit of the quasi-DA.

	Our evolved results for the twist-2 (axialvector)  and twist-3 (tensor) pion and kaon DA amplitudes,   are very  close to the DSE results, and consistent with
	the recently reported lattice results for the twist-2.  Our evolved results for the twist-3 (pseudoscalar) for the pion and kaon DA amplitudes are different from those
	obtained using the DSE, but very close to the QCD asymptotic results. It is rather remarkable, that our pion and kaon DA amplitudes probe  specifically the running 
	emergent topological quark mass  in the dilute RIV, which is the dominant component of the QCD vacuum at low resolution.

	\vskip 1cm
	{\bf Acknowledgements}
	
	This work is supported by the Office of Science, U.S. Department of Energy under Contract No. DE-FG-88ER40388.

\section{Appendices} 

\subsection{Details of Eq.~\ref{qpda3}}
~\label{appendixA}

We start from (\ref{PHIP}) and perform the analytical continuation $k_4\rightarrow ik_4$, followed by the shift  $k_4\rightarrow k_4+(x-\frac{1}{2})P_z$.
The result is 
	
	\begin{equation}\label{qpda2}
		\tilde \phi^P_0(x,P_z)=
		\frac{2iN_cP_z}{f_\pi^2\chi_{\pi}}\int \frac{d^2k_\perp (idk_4)}{(2\pi)^4} \left(M(y_1)M(y_2)\right)^{1/2}\frac{y_1^2+y_2^2-m_\pi^2}{(y_1^2+M_1^2)(y_2^2+M_2^2)}
	\end{equation}
	where
	
	\begin{align}\label{kinematics}
		\begin{split}
		y_1^\mu &=\left(\vec{k}_{\perp},xP_z,i(k_4+xE)\right)  \\
		y_1^2&=-k_4(k_4+2xE)+k_\perp^2-x^2m_{\pi}^2 - i\epsilon  \\
		y_1^2+M_1^2 &=-(k_4-k_{4+})(k_4-k_{4-})  \\
		&\\
		y_2^\mu &=\left(\vec{k}_{\perp},-\bar x P_z,i(k_4-\bar x E)\right) \\
		y_2^2&=-k_4(k_4-2\bar x E)+k_\perp^2-\bar x^2 m_{\pi}^2 - i\epsilon \\
		 y_2^2+M_2^2&=-(k_4-\bar{k}_{4+})(k_4-\bar{k}_{4-})
		\end{split}
	\end{align}
The emergent constituent quark mass $M(y)$ is characterized by branch-points.
	In~\cite{Kock:2020} we have shown that the analysis retaining the branch points in $M(y)$ 
	is similar to the one following from  the modified effective mass at large $P_z$, 
	
	\begin{equation}\label{EffectiveCutoff}
		M(y)\rightarrow M\left(\frac{k_{\perp}}{\lambda \sqrt{|x\bar x |}}\right)
	\end{equation}
This removes the explicit $k_4$-dependence in $M(y)$, so we can evaluate the $k_4$ integral by residues.
	The value of $\lambda$ is then fixed to reproduce unit normalization of the DA. Note that in~\cite{Kock:2020}, 
	we used the additional constraint $k_\perp>M(0)$ in the cutoff, which does not affect the power counting, but is not necessary.
	
	%the substitution (\ref{EffectiveCutoff}) was slightly different. Due to a subtle difference in integration range, the substitution in that paper was given by 
	
	%\begin{equation}\label{EffectiveCutoff2}
	%	M(y)\rightarrow M\left(\frac{\sqrt{k_{\perp}^2+M(0)^2}}{\lambda \sqrt{|x\bar x|}}\right)
	%\end{equation}

	%There is a minor ambiguity in which should be correct, as both are identical to leading order in $\alpha$. This is discussed further in appendix \hyperref[appendixA]{A}.
	
	With the above in mind, the integrand in (\ref{qpda2}) has 4 poles in the complex $k_4$ plane, $\{k_{4\pm},\bar{k}_{\pm}\}$. For  large $P_z$, two poles $\{k_{4+},\bar k_{4-}\}$ tend toward the origin, whereas the other two $\{k_{4-},\bar k_{4+}\}$ tend toward infinity.
	
	\begin{align}
		k_{4+}&\approx \frac{k_\perp^2+M_1^2-x^2m_{\pi}^2-i\epsilon}{2xE} &  \bar k_{4-}&\approx \frac{k_\perp^2+M_2^2-\bar x^2 m_{\pi}^2-i\epsilon}{-2\bar x E} 	\notag \\
		k_{4-}&\approx -2xE +i\epsilon & \bar k_{4+} &\approx 2\bar x E - i\epsilon
	\end{align}
Notice that only for the physical domain $x\bar x > 0$ do the poles close to the origin appear on each half-plane. In the unphysical domain $x\bar x <0$, both of these poles lie in the same half-plane, which after closing the contour in the opposite half-plane gives a vanishing result. We encapsulate this fact, that the leading-order PDAs have support only in the physical domain $x\bar x >0$, with on overal $\theta (x\bar x )$. 
	We close the contour for (\ref{qpda2}) in the UHP, picking up only $\bar k_{4-}$ and $k_{4-}$. At the location of the first pole $\bar k_{4-}$, we have
	
	\begin{align}
		y_2^2+M_2^2&=0 & y_1^2&\approx \frac{1}{\bar x}\left[k_{\perp}^2+M_2^2-x\bar x m_{\pi}^2\right]-M_2^2 & \bar k_{4-}-\bar k_{4+}&=-2\bar x E
	\end{align}
At the second pole $k_{4-}$, we have 
	
	\begin{align}
		y_1^2+M_1^2&=0  & y_2^2&\approx -4E^2 x &  k_{4-}- k_{4+}&=-2\bar x E
	\end{align}

\subsection{Normalization $\lambda$ Values}
\label{appendixB}

\begin{table}[ht]
	\centering
	\begin{tabular}{ |P{0.7cm}|P{2.7cm}|P{2.7cm}|P{2.7cm}| } 
	\hline
	 & original & mod(1) & mod(2)  \\
	\hline
	$\lambda^A_\pi$ & 4.918 & 4.906 & 5.839 \\
	$\lambda^A_K$ & 5.944 & 6.125 & 6.877 \\
	$\lambda^P_\pi$ & 1.631 & 1.633 & 1.906 \\
	$\lambda^P_K$ & 1.556 & 1.595 & 1.846 \\
	$\lambda^T_\pi$ & 1.5337 & 1.538 & 1.988  \\
	$\lambda^T_K$ & 1.466 & 1.594 &  1.932  \\
	\hline
\end{tabular}
\caption{Values of $\lambda$ which normalize the DAs. A,P, and T correspond to axial-vector (twist-2), pseudoscalar (twist-3), and pseudotensor (twist-3) DAs respectively. "original" corresponds to our main expressions for the DA, (\ref{qpda3}), (\ref{qpda5}), and (\ref{twist2}). The modifications "mod(1)" and "mod(2)" are discussed in appendix \hyperref[appendixC]{C}.}
\label{lambda_table}
\end{table}

\begin{figure}
	\centering
	\subfloat{\label{fig_PDA_MassChanges:a}\includegraphics[width=.49\textwidth,height=0.32\textwidth]{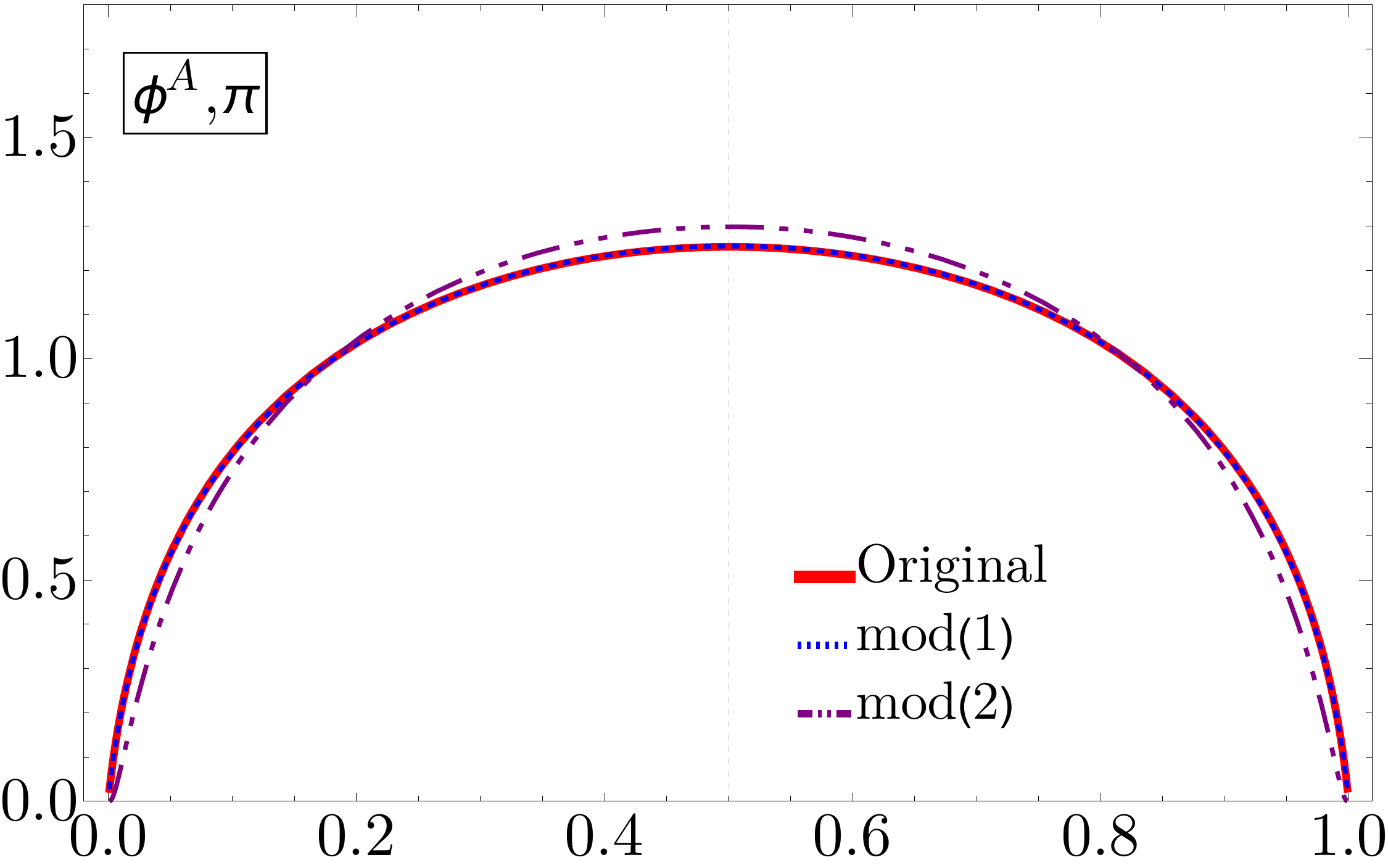}}\hfill
	\subfloat{\label{fig_PDA_MassChanges:b}\includegraphics[width=.49\textwidth,height=0.32\textwidth]{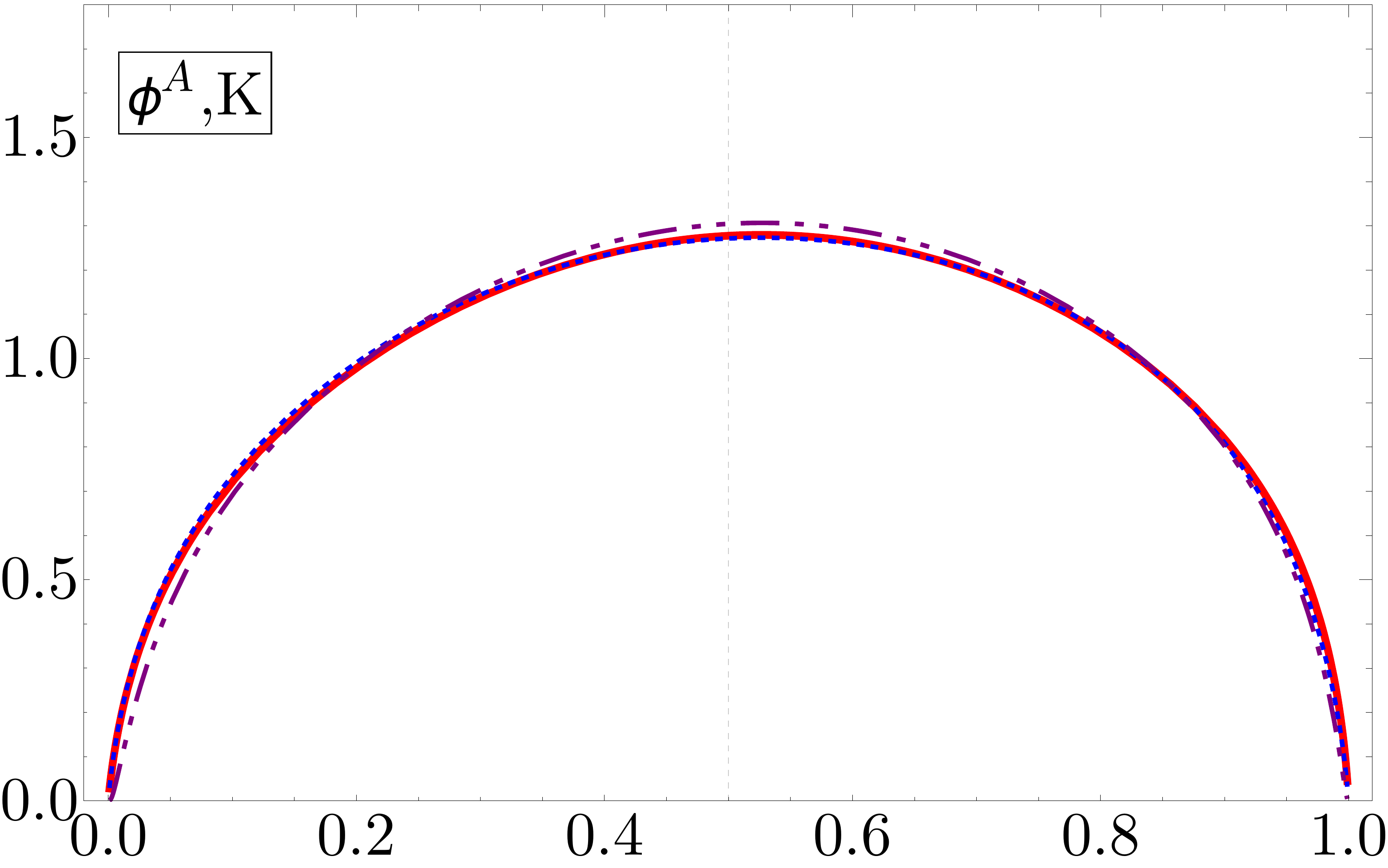}}\hfill
	\subfloat{\label{fig_PDA_MassChanges:c}\includegraphics[width=.49\textwidth,height=0.32\textwidth]{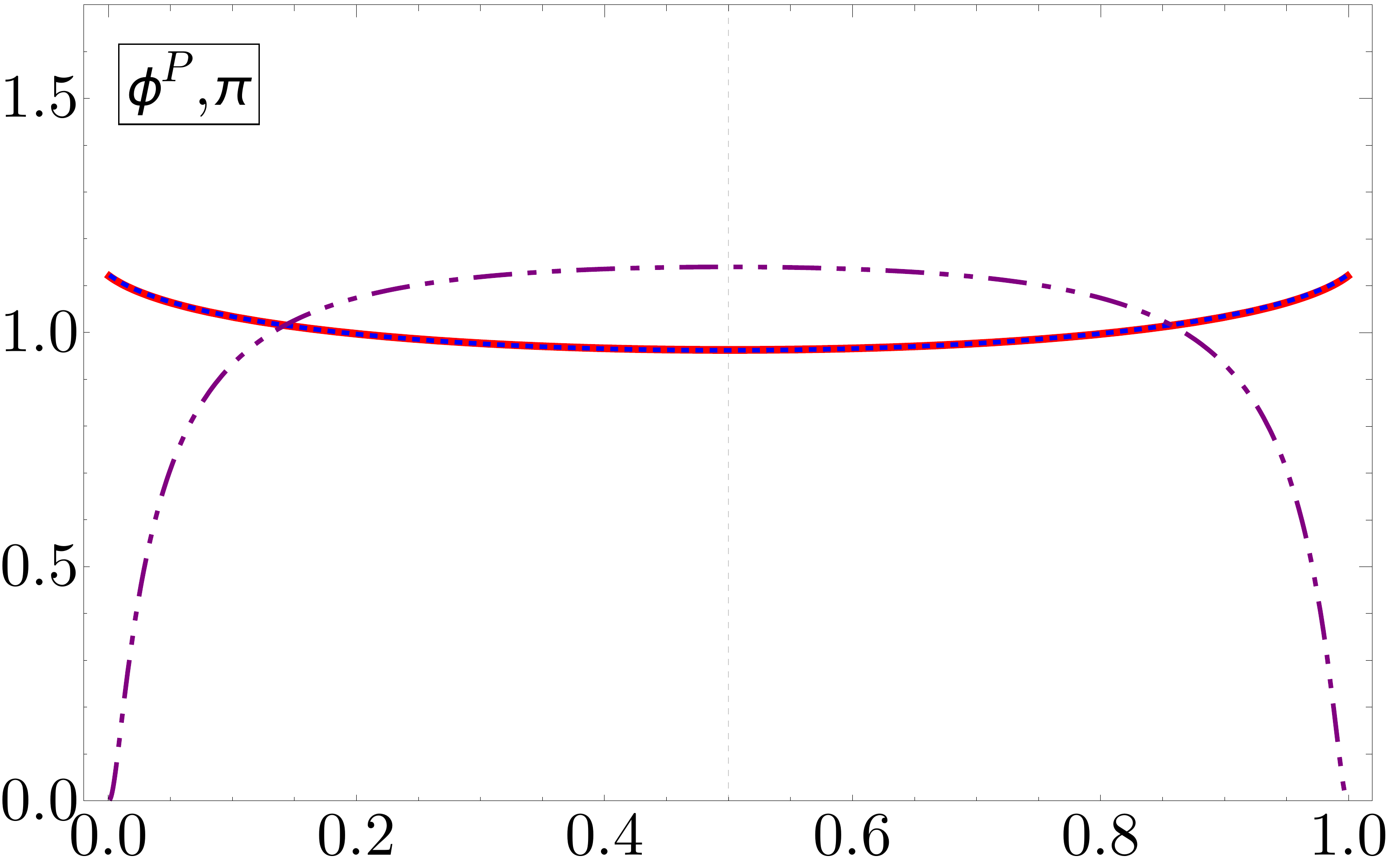}}\hfill
	\subfloat{\label{fig_PDA_MassChanges:d}\includegraphics[width=.49\textwidth,height=0.32\textwidth]{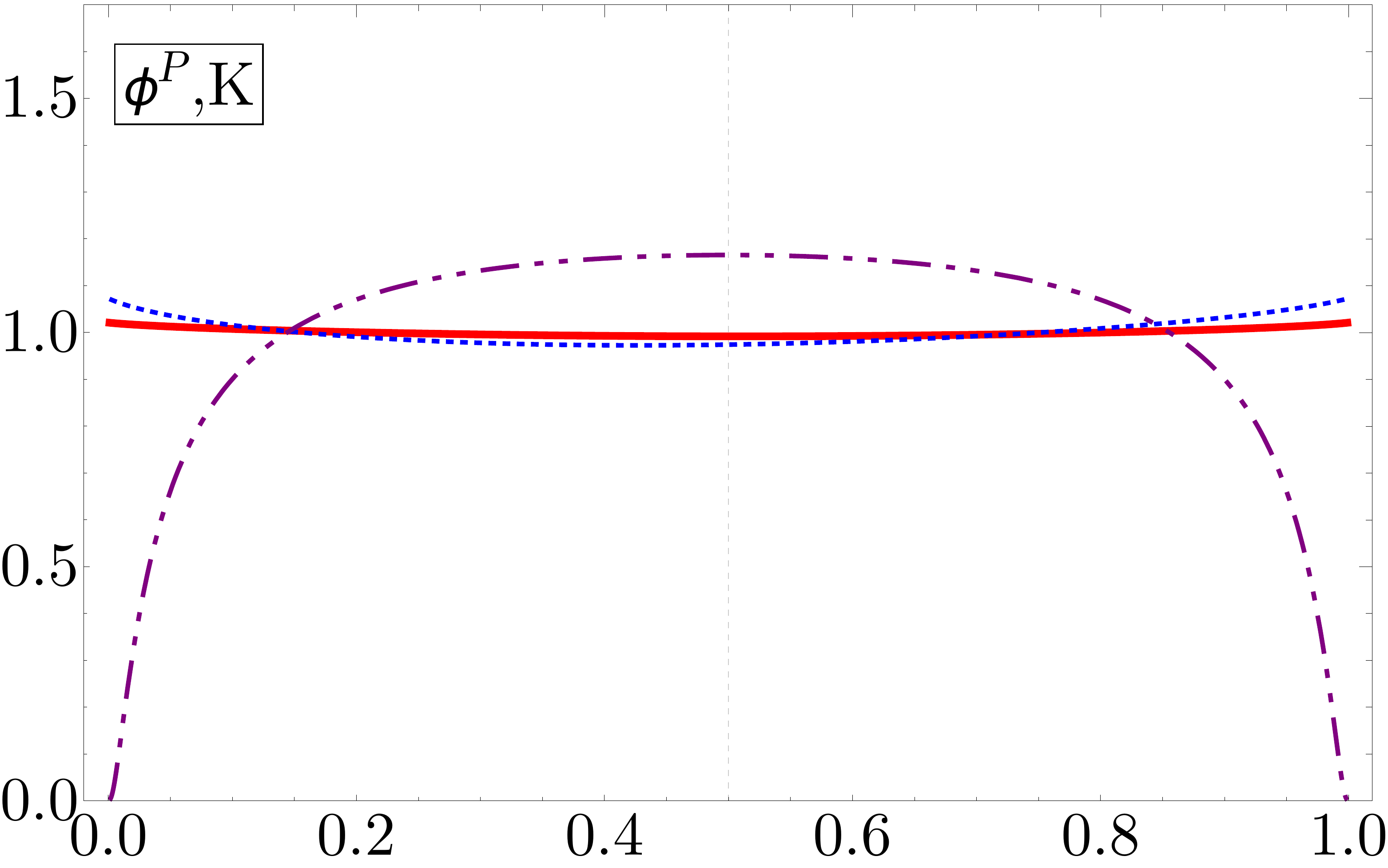}}\hfill
	\subfloat{\label{fig_PDA_MassChanges:e}\includegraphics[width=.49\textwidth,height=0.32\textwidth]{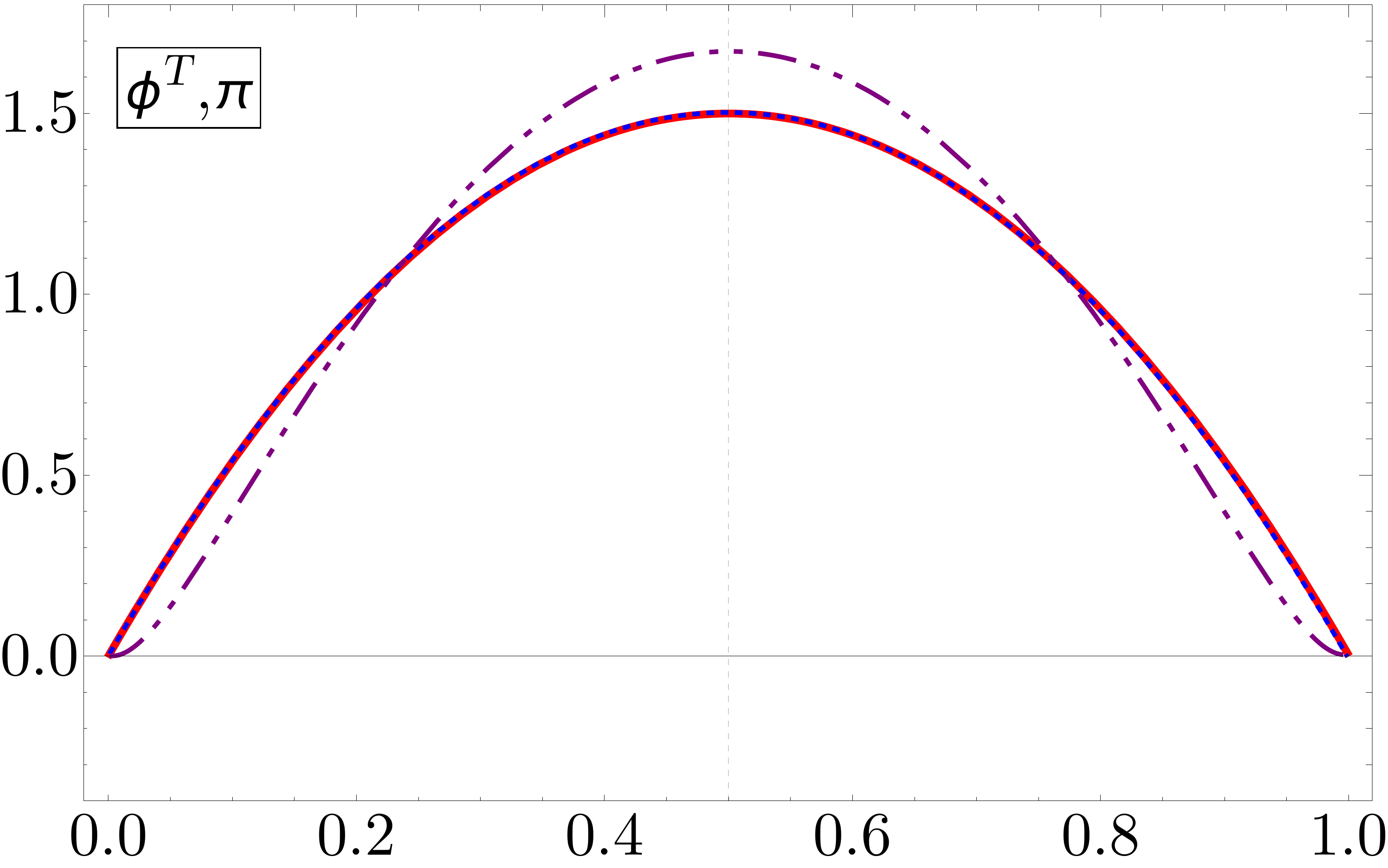}}\hfill
	\subfloat{\label{fig_PDA_MassChanges:f}\includegraphics[width=.49\textwidth,height=0.32\textwidth]{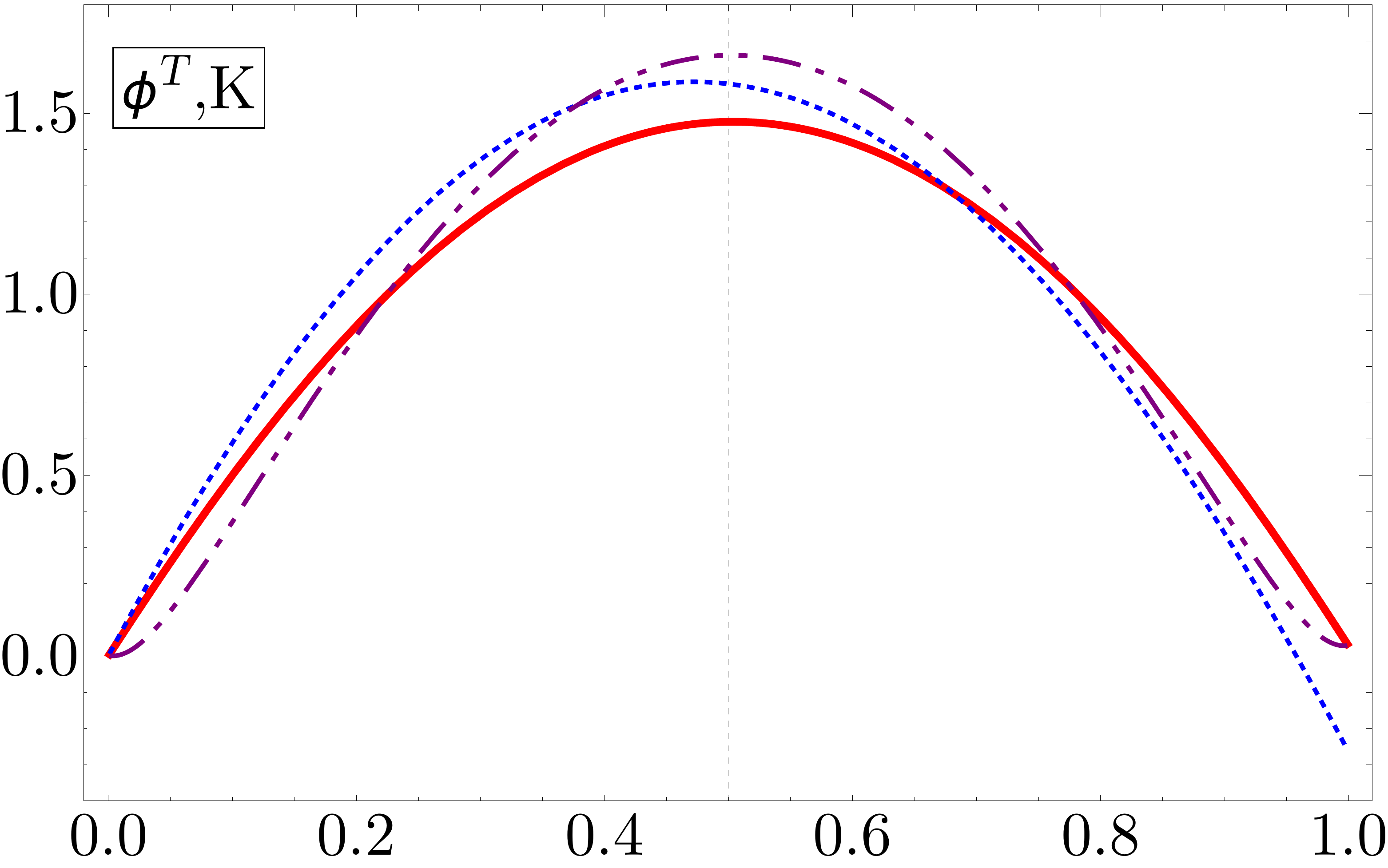}}\hfill
	\caption{Unevolved DA, at $Q=Q_0= 631\,\textrm{MeV}$, with certain modifications to the effective mass. "Original" (red-thick-solid) denotes 
	(\ref{qpda3}), (\ref{qpda5}), and (\ref{twist2}). "mod(1)" (blue-dotted) denotes the shifted effective mass, and  "mod(2)" (purple-dash-dot-dot) 
	the shifted cutoff. See text. }
	\label{fig_PDA_MassChanges}
\end{figure}

\subsection{Modifications to the Constituent Quark Mass}
\label{appendixC}

Our effective quark mass obtained at leading order in $\alpha$ is given by (\ref{ConstituentMass}). A plot of the zero-momentum limit as a function of current mass $m$ is shown in fig. \ref{fig_MASS}.
This constituent quark mass has the interesting property of being approximately constant for $0\,\textrm{MeV}<m<180\,\textrm{MeV}$, assuming phenomenological values of $\rho$ and $\alpha$. 
%Moreover, $M(0,m_s)<M(0,m_u)$, in contradiction with the common assumption that $M(0,m)\approx M(0)+m\implies M(0,m_s)>M(0,m_u)$.
Although obtained from a $O(\alpha )$ integral equation, (\ref{ConstituentMass}) contains higher powers of $\alpha$ resummed into dependence on $m$ through $\xi$. If we maintain strict power counting in $\alpha$ and simultaneously assume sufficiently small quark mass $m$ such that $\xi\ll 1$ (for our parameter values this is not true), then we arrive at the approximation $M(k,m)\approx M(k)+m$. We insert this approximation into the expressions for the DAs and see how they change. In fig. \ref{fig_PDA_MassChanges}, this change is denoted by "mod(1)". The most notable resulting change is seen in the kaon pseudotensor DA, which no longer approaches zero at $x\rightarrow 1$. A slight restoration of $x\leftrightarrow \bar x$ symmetry is seen in the kaon twist-2 DA. All other DAs are relatively unchanged.

Finally, we consider the effect of restricting $k_\perp >M(0,m)$ in the cutoff (\ref{EffectiveCutoff}),
as was noted in  section~\ref{QDA-3}.
%\begin{equation}\label{ModEffectiveCutoff}
%	M(y)\rightarrow M\left(\frac{\sqrt{k_\perp^2+M^2(0)}}{\lambda \sqrt{|x\bar x|}}\right)
%\end{equation}
This modification was implicitly present in our previous paper~\cite{Kock:2020}.
%(see the discussion at the beginning of section B). 
In fig. \ref{fig_PDA_MassChanges} this change is denoted by "mod(2)". This results in a slight narrowing of the axial-vector (twist-2) and pseudotensor (twist-3) DAs. The biggest change is seen in the pseudoscalar (twist-3) DA, which becomes concave.

\bibliography{twist3}

\end{document}